\documentclass[11pt] {article}
\usepackage{epsfig}
\catcode`@=11

\def\@citex[#1]#2{\if@filesw\immediate\write\@auxout{\string\citation{#2}}\fi
  \def\@citea{}\@cite{\@for\@citeb:=#2\do
    {\@citea\def\@citea{,\penalty\@m}\@ifundefined
      {b@\@citeb}{{\bf ?}\@warning
       {Citation `\@citeb' on page \thepage \space undefined}}%
\hbox{\csname b@\@citeb\endcsname}}}{#1}}

\def\citer{\@ifnextchar [{\@tempswatrue\@citexr}{\@tempswafalse\@citexr[]}}

\def\@citexr[#1]#2{\if@filesw\immediate\write\@auxout{\string\citation{#2}}\fi
  \def\@citea{}\@cite{\@for\@citeb:=#2\do
    {\@citea\def\@citea{--\penalty\@m}\@ifundefined
       {b@\@citeb}{{\bf ?}\@warning
       {Citation `\@citeb' on page \thepage \space undefined}}%
\hbox{\csname b@\@citeb\endcsname}}}{#1}}
\catcode`@=12
\def\s{\hat{s}}

\def\ms{\hat{m}_s}

\def\mc{\hat{m}_c}

\def\q{\hat{q}}
\def\bxsll{$B \rightarrow X_s \ell^+ \ell^- $}

\title{  { \bf
Phenomenological Profiles of the Inclusive Hadron Spectra in the Decay
 \bxsll  }}
\author{\vspace{1cm}\\
   {\bf A.~Ali\thanks{E-mail address: ali@x4u2.desy.de} ~and 
G.~Hiller\thanks{E-mail address: ghiller@x4u2.desy.de}}\\
         Deutsches Elektronen-Synchrotron DESY, Hamburg \\
        \vspace{5mm}\\}
\date{}

\begin{document}
\setlength{\baselineskip}{24pt}

\maketitle
\begin{picture}(0,0)
       \put(325,310){DESY 98-031}
       \put(325,295){hep-ph/9807yyy}
       \put(325,280){July 1998}
\end{picture}
\vspace{-24pt}
\setlength{\baselineskip}{7mm}

\begin{abstract}
Hadron spectra and hadronic moments in the decay $B \to X_s \ell^+ \ell^-$
are calculated taking into account both the short-distance and long-distance
contributions in the decay amplitude using a Fermi motion (FM) model
to incorporate the $B$-meson wave-function effects. The measured
branching ratios for the inclusive decays
$B \to X_s+ (J/\psi,\psi^\prime,...)\to X_s \ell^+ \ell^-$ are used
to fix the normalization of the long-distance contribution.
The momentum distribution of the $J/\psi$ measured by the CLEO collaboration
is fitted in the FM model which is then used to calculate the hadronic spectra
from the resonant contribution also away from the $J/\psi$-resonance.
We also study the effect of various descriptions of the resonant and
non-resonant $c\bar{c}$ contributions in $B \to X_s \ell^+ 
\ell^-$ existing in the literature on the hadron energy and invariant
mass spectra, and in the Forward-Backward asymmetry. Selective cuts on the
hadron and dilepton invariant masses can be used to reduce the 
$B\bar{B}$ background and resonant contribution and, as an example, we work 
out the hadron spectra with the
experimental cuts used by the CLEO collaboration in searching for the decay
$B \to X_s \ell^+ \ell^-$. We show that data from the forthcoming B
facilities could be used effectively to measure the short-distance
contribution in $B \to X_s \ell^+ \ell^-$, enabling precise determination of
the FM model and heavy quark effective theory parameters $\lambda_1$ and
$\bar{\Lambda}$.

\vspace*{1.5cm}
\centerline{(Submitted to Physical Review D)}

\end{abstract}

\thispagestyle{empty}
\newpage
\setcounter{page}{1}

\section{Introduction}

 In two earlier papers \cite{AH98-1,AH98-2}, we have worked out the
short-distance (SD) contribution to the hadron
energy and hadronic invariant  mass spectra and the first two spectral 
moments in the inclusive decay $B \to X_s \ell^+ \ell^-$. The
calculations reported in these papers were based on the leading order 
perturbative QCD corrections in
$\alpha_s$ and leading power corrections in $1/m_b^2$ using the heavy
quark expansion technique (HQET) \cite{georgi,MW}. In particular, we 
worked out the dependence of the moments $\langle S_H^n\rangle$ and 
$\langle E_H^n \rangle$, for $n=1,2$,
valid up to ${\cal O}(\alpha_s/m_B^2,1/m_B^3)$. It was argued that their 
measurements in forthcoming experiments could be combined with the 
improved measurements of the same in the semileptoic decays $B \to X \ell 
\nu_\ell$ to determine the HQET parameters $\lambda_1$ and $\bar{\Lambda}$
precisely. Since these parameters are endemic to most applications of HQET,
their precise determination would reduce the present theoretical 
uncertainties improving the standard model calculations. In particular, the 
determinations of the CKM matrix elements $V_{ub}$ and $V_{cb}$ would be
considerably improved. 
The correlations resulting from (assumed) values of $\langle S_H \rangle$
and $\langle S_H^2 \rangle$ were shown and compared with the constraints 
emerging from the analysis of the decay $B \to X \ell \nu_\ell$ reported in
\cite{gremm}.  The hadron spectra and spectral  
moments were also calculated in a phenomenological Fermi motion model (FM)
\cite{aliqcd}. In particular, the remarkable similarity of the hadronic 
moments in the HQET and FM model approaches was quantified.
Finally, the effects of the experimental cuts on the spectra used in the 
searches for the
decay $B \to X_s \ell^+ \ell^-$ by the CLEO collaboration \cite{cleobsll97}
were studied. However, the effects of the long-distance contributions
were not included in the hadron spectra or the spectral moments.

 The aim of this paper is to calculate the profile 
of hadron energy and invariant mass spectra by   
incorporating the effects of the
long-distance (LD) contributions in the decay $B \to X_s \ell^+ \ell^-$.
As opposed to the SD-contribution discussed in \cite{AH98-1,AH98-2}, the 
LD-contributions are estimated phenomenologically. To that end,  
the branching ratios for $B \to (J/\psi,\psi^\prime,...)
+ X_s \to X_s \ell^+ \ell^-$ are described in the factorization approach 
\citer{BSW87,BHP96}, with the  data
fixing the normalization and phase of the LD-contribution in \bxsll at the
$J/\psi,\psi^\prime,...$ resonances \cite{AHHM97}.
Assuming  a Breit-Wigner form,  one can extrapolate the 
dilepton mass spectra away from the $J/\psi,\psi^\prime,...$ resonances.
Using this and the SD-contribution, various distributions in \bxsll
have been worked out in the literature \citer{AHHM97,MKS98}.
We use the FM model to incorporate the wave-function effects. In the process
of doing this, we also show that the FM model \cite{aliqcd} provides an
adequate description of the $J/\psi$-momentum distribution in the decay
$B \to J/\psi + X_s \to X_s \ell^+ \ell^-$
measured by the CLEO collaboration \cite{CLEOjpsi94}. This confirms a
similar and earlier study  on this point
\cite{PPS97}. However, we have redone
a fit of the CLEO-data on the $J/\psi$-momentum spectrum and prefer
somewhat different FM model
parameters than the ones presented in \cite{PPS97}, motivated by the
analysis of the photon energy spectrum in the decay $B \to X_s + \gamma$
\cite{ag95}
and theoretical consideration on the $b$-quark mass \cite{neubertsachr}.

The prescription of adding the SD and 
LD-contributions in the decay \bxsll is not unique, which introduces a 
theoretical dispersion in the resulting spectra. Related to this is the
inherent uncertainty concerning the extrapolation of the
resonant part far away from the resonances \cite{amm91,ahmady96,KS96}.
We study these uncertainties by working out a number of Ans\"atze
used in the literature. The
dispersion on the spectra emerging from various approaches can then be taken
as a measure of theoretical systematic errors from these sources. We
analyze the dilepton invariant mass distribution, the Forward Backward (FB) 
asymmetry, hadron energy and hadronic invariant mass spectra and the hadron 
spectral moments in this context. 
The LD/SD-related uncertainties are found to be small for the
hadron spectra and spectral moments. Some of these issues were also 
discussed in the context of exclusive decays $B \to (K,K^*) \ell^+
\ell^-$ in the second reference cited in \cite{MKS98}.
  
  Finally, we study the effect of experimental cuts on the branching ratios,
hadron spectra and spectral moments in \bxsll. For that purpose we take the
cuts used by the CLEO collaboration \cite{cleobsll97}, which involve
dilepton and hadronic invariant masses. We present a comparative analysis
of these cuts on the inclusive hadron energy and invariant hadronic mass
distributions with and without the $c\bar{c}$-resonant contributions.
This shows that the cuts employed in \cite{cleobsll97} are effective in
removing the resonant part and given data one could study the more 
interesting SD-contribution in \bxsll.

This paper is organized as follows: In section 2, we define the kinematics 
of the process $B \to X_s \ell^+ \ell^-$ and introduce the quantities of
dynamical interest in the framework of an effective Hamiltonian. 
Section 3 describes the wave-function effects in the FM model
\cite{aliqcd} in the hadron energy and hadronic invariant mass 
spectra. The resulting inclusive hadron spectra in \bxsll , including the 
long-distance effects in terms of the $J/\psi,\psi^\prime,...$ resonances,
are presented. Comparison with the $J/\psi$-momentum spectrum
measured by the CLEO collaboration is shown,
constraining the FM model parameters. Theoretical uncertainties
in the effective coefficients ${\rm Re} ~C_9^{\rm eff}$ and $\vert 
C_9^{\rm eff} \vert$ from the various prescriptions are displayed here
and the resulting spectra in hadron energy, dilepton invariant mass and 
the FB-asymmetry are presented.
Hadronic spectral moments are calculated in the FM model taking into account 
the LD contributions. Comparison with the corresponding quantities derived
from the SD-contribution alone using the HQET and FM 
models in \cite{AH98-1,AH98-2} is 
also presented. In section 4, the effects of the 
experimental cuts used in the CLEO
analysis of \bxsll are studied  and the resulting spectra are  
presented in terms of several figures.
 Estimates of the
branching ratios ${\cal B}(B \to X_s \ell^+ \ell^-)$ for $\ell =e,\mu$
are also presented here, together with estimates of the survival
probability for the CLEO cuts, using the FM model.
Section 5 contains a summary of our work and some 
concluding remarks. 

\section{\bf The Decay \bxsll in the Effective Hamiltonian Approach}

\subsection{Kinematics}

The kinematics for the decay in question at the partonic and hadronic level
are defined in \cite{AH98-1,AH98-2}. Hence, we will be short here. 
The parton level kinematics is given by 
\begin{equation}
b (p_b) \to s (p_s) (+g (p_g))+\ell^{+} (p_{+})+\ell^{-}(p_{-}) ~,
\end{equation} 
where $g$ denotes a gluon from the $O(\alpha_s)$ correction. The
corresponding kinematics at the hadron level is defined as:
\begin{equation}
B (p_B) \to X_s (p_H)+\ell^+ (p_{+})+\ell^{-} (p_{-})~.
\end{equation}
We define the momentum transfer to the lepton pair and the invariant 
mass of the dilepton system, respectively, as
\begin{eqnarray}
q &\equiv & p_{+}+p_{-} \; , \\
s &\equiv & q^2 \; .
\end{eqnarray}
Further, we define a 4-vector $v$, which denotes the velocity of both the
$b$-quark and the $B$-meson, $p_b=m_b v$ and $p_B=m_B v$.
The hadronic invariant mass is denoted by  $S_H \equiv p_H^2$ and 
$E_H$ denotes the hadron energy in the final state. 
The corresponding quantities at parton level are the 
invariant mass $s_0$ and the scaled parton energy $x_0\equiv \frac{E_0}{m_b}$.
In parton model without gluon bremsstrahlung, this simplifies to
$s_0=m_s^2$ and $x_0$ becomes directly related to the dilepton invariant mass 
$x_0=1/2(1-\s +\ms^2)$.
Here and in what follows, the dimensionless variables with a
hat are related to the dimensionful variables by the scale $m_b$,
the $b$-quark mass, e.g., $\s= \frac{s}{m_b^2}$, $\ms=\frac{m_s}{m_b}$ etc.
{}From momentum conservation the following equalities hold in the $b$-quark,
equivalently $B$-meson, rest frame ($v=(1,0,0,0)$):
\begin{eqnarray}
x_0  &=& 1- v \cdot \q \, \, ,\nonumber \\
\s_0 &=& 1 -2 v \cdot \q + \s \, \, ,\label{eq:kin} \\
E_H &=& m_B-v \cdot q \, \, ,\nonumber \\
S_H &=& m_B^2 -2 m_B v \cdot q  + s \, \, .
\end{eqnarray}
The relation between the kinematic variables of the parton model and the 
hadronic states can be seen in \cite{AH98-1,AH98-2}. 

\subsection{Matrix element for the decay \bxsll}

The effective 
Hamiltonian obtained by integrating out the top quark and the $W^\pm$ bosons
is given as
\begin{eqnarray}\label{heffbsll}
{\cal H}_{eff}(b \to s + X, \, X=\gamma, \, \ell^{+} \ell^{-}) 
= - \frac{4 G_F}{\sqrt{2}} V_{ts}^* V_{tb}
 \left[ \sum_{i=1}^{6} C_i (\mu)  O_i 
+ C_7 (\mu) \frac{e}{16 \pi^2}
          \bar{s}_{\alpha} \sigma_{\mu \nu} (m_b R + m_s L) b_{\alpha}
                F^{\mu \nu} 
\right. \nonumber \\
 \left.
+C_8 (\mu) O_8 
+ C_9 (\mu) \frac{e^2}{16 \pi^2}\bar{s}_\alpha \gamma^{\mu} L b_\alpha
\bar{\ell} \gamma_{\mu} \ell 
+ C_{10}  \frac{e^2}{16 \pi^2} \bar{s}_\alpha \gamma^{\mu} L
b_\alpha \bar{\ell} \gamma_{\mu}\gamma_5 \ell \right] \, ,
\end{eqnarray}
where $L$ and $R$ denote chiral projections, $L(R)=1/2(1\mp \gamma_5)$,
 $V_{ij}$ are the CKM matrix elements and the
CKM unitarity has been used in factoring out the product $V_{ts}^\ast
V_{tb}$. The operator basis is taken from \cite{AHHM97}, where also the 
Four-Fermi operators $O_{1},\dots ,O_{6}$ and the chromomagnetic 
operator $O_8$ can be seen.
Note that $O_8$ does not contribute to the decay \bxsll in the 
approximation which we use here. The $C_i(\mu)$ are the Wilson coefficients,
which depend, in general, on the renormalization scale $\mu$,
except for $C_{10}$. 

The matrix element for the decay \bxsll can be factorized 
into a leptonic and a hadronic part as
\begin{equation}
        {\cal M (\mbox{\bxsll})} =
        \frac{G_F \alpha}{\sqrt{2} \pi} \, V_{ts}^\ast V_{tb} \, 
        \left( {\Gamma^L}_\mu \, {L^L}^\mu 
        +  {\Gamma^R}_\mu \, {L^R}^\mu \right) \, ,
\end{equation}
with
\begin{eqnarray}
        {L^{L/R}}_\mu & \equiv & 
                \bar{\ell} \, \gamma_\mu \, L(R) \, \ell \, , \\
        {\Gamma^{L/R}}_\mu & \equiv & 
                \bar{s} \left[ 
                R \, \gamma_\mu 
                        \left( C_9^{\mbox{eff}}(\s) \mp C_{10} 
                          + 2 C_7^{\mbox{eff}} \, 
                        \frac{\hat{\not{q}}}{\s} \right)
                + 2 \hat{m}_s \, C_7^{\mbox{eff}} \, \gamma_\mu \, 
                        \frac{\hat{\not{q}}}{\s} L 
                \right] b \, .  
        \label{eqn:gammai}
\end{eqnarray}
The effective Wilson coefficient $C_9^{\mbox{eff}}(\s)$ receives 
contributions from various pieces. Since the
resonant $c\bar{c}$ states also contribute to $C_9^{\mbox{eff}}(\s)$,
the contribution given below is just the perturbative part:
\begin{eqnarray}
C_9^{\mbox{eff}}(\s)|_{pert}=C_9 \eta(\s) + Y(\s) \, .
\end{eqnarray}
Here $\eta(\s)$ and $Y(\s)$ represent the ${\cal{O}}(\alpha_s)$ correction 
\cite{jezkuhn}
and the one loop matrix element of the 
Four-Fermi operators \cite{burasmuenz,misiakE}, respectively. 
While $C_9$ is a renormalization scheme-dependent quantity, this
dependence cancels out with the corresponding one in the function $Y(\s)$
(the value of $\xi$, see below).
To be self-contained, we list
the two functions in $C_9^{\mbox{eff}}(\s)$:
\begin{eqnarray}
\label{Ypert}
        Y(\s) & = & g(\mc,\s)
                \left(3 \, C_1 + C_2 + 3 \, C_3
                + C_4 + 3 \, C_5 + C_6 \right)
\nonumber \\
        & & - \frac{1}{2} g(1,\s)
                \left( 4 \, C_3 + 4 \, C_4 + 3 \,
                C_5 + C_6 \right)
         - \frac{1}{2} g(0,\s) \left( C_3 +
                3 \, C_4 \right) \nonumber \\
        & &     + \frac{2}{9} \left( 3 \, C_3 + C_4 +
                3 \, C_5 + C_6 \right)
             - \xi \, \frac{4}{9} \left( 3 \, C_1 +
                C_2 - C_3 - 3 \, C_4 \right),
                \label{eqn:y} \\
        \eta(\s) & = & 1 + \frac{\alpha_s(\mu)}{\pi}
                \omega(\s) ~,
\end{eqnarray}
 \begin{equation}
        \xi = \left\{
                \begin{array}{ll}
                        0       & \mbox{(NDR)}, \\   
                        -1      & \mbox{(HV)},
                \end{array}
                \right.
\end{equation}

\begin{eqnarray}
\label{gpert}   
g(z,\hat{s}) &=& -\frac{8}{9}\ln (\frac{m_b}{\mu})
 -\frac{8}{9} \ln z + \frac{8}{27} +\frac{4}{9}y
-\frac{2}{9}(2 + y) \sqrt{\vert 1-y \vert}\nonumber\\
&\times & \left[\Theta(1-y)(\ln\frac{1+\sqrt{1-y}}{1-\sqrt{1-y}} -i\pi )
+\Theta(y-1) 2 \arctan \frac{1}{\sqrt{y-1}} \right] ~, \\
g(0,\hat{s})& =& \frac{8}{27}-\frac{8}{9}\ln (\frac{m_b}{\mu})
              -\frac{4}{9}\ln \hat{s} + \frac{4}{9}i\pi ~,
\end{eqnarray}
where $y=4z^2/\hat{s}$, and
\begin{eqnarray}
\omega(\hat{s}) &=& -\frac{2}{9}\pi^2 -\frac{4}{3}{\mbox Li}_2(\hat{s})
-\frac{2}{3}
\ln \hat{s} \ln(1-\hat{s}) -
\frac{5+4\hat{s}}{3(1+2\hat{s})}\ln(1-\hat{s})\nonumber\\
&-& \frac{2\hat{s}(1+\hat{s})(1-2\hat{s})}{3(1-\hat{s})^2(1+2\hat{s})}
\ln \hat{s} + \frac{5 + 9\hat{s} -6\hat{s}^2}{6(1-\hat{s})(1+2 \hat{s})}~.
\label{omegahats}
\end{eqnarray}
Above, (NDR) and (HV) correspond to the naive dimensional regularization
and the 't Hooft-Veltman schemes, respectively. Note that the function
$g(\mc,\hat{s})(3C_1 + C_2+...)$ given above is the perturbative
contribution to the effective coefficient $C_9^{\mbox{eff}}(\hat{s})$
from the $c\bar{c}$ loop, to which we have referred in the introduction
and to whose discussion we shall return in section 3.
The other Wilson coefficients in leading logarithmic approximation
can be seen in \cite{burasmuenz}.

\begin{table}[h]
        \begin{center}
        \begin{tabular}{|l|l|}
        \hline
        \multicolumn{1}{|c|}{Parameter} & 
                \multicolumn{1}{|c|}{Value}     \\
        \hline \hline
        $m_W$                   & $80.26$ (GeV) \\
        $m_Z$                   & $91.19$ (GeV) \\
        $\sin^2 \theta_W $      & $0.2325$ \\
        $m_s$                   & $0.2$ (GeV)   \\
        $m_c$                   & $1.4$ (GeV) \\
        $m_b$                   & $4.8$ (GeV) \\
        $m_t$                   & $175 \pm 5$ (GeV)     \\
        $\mu$                   & ${m_{b}}^{+m_{b}}_{-m_{b}/2}$        \\
        $\Lambda_{QCD}^{(5)}$   & $0.214^{+0.066}_{-0.054}$ (GeV)       \\
        $\alpha^{-1}$     & 129           \\
        $\alpha_s (m_Z) $       & $0.117 \pm 0.005$ \\
        ${\cal B}_{sl}$         & $(10.4 \pm 0.4)$ \%   \\
        $\lambda_1$             & $-0.20$ (GeV$^2$) \\   
        $\lambda_2$             & $+0.12$ (GeV$^2$) \\
        \hline
        \end{tabular}
        \end{center}
\caption{\it Default values of the input parameters and errors used in the 
numerical calculations.}
\label{parameters}
\end{table}
\begin{table}[h]
        \begin{center}
        \begin{tabular}{|c|c|c|c|c|c|c|c|c|c|}
        \hline
        \multicolumn{1}{|c|}{ $C_1$}       & 
        \multicolumn{1}{|c|}{ $C_2$}       &
        \multicolumn{1}{|c|}{ $C_3$}       & 
        \multicolumn{1}{|c|}{ $C_4$}       &
        \multicolumn{1}{|c|}{ $C_5$}       & 
        \multicolumn{1}{|c|}{ $C_6$}       &
        \multicolumn{1}{|c|}{ $C_7^{\mbox{eff}}$}       & 
        \multicolumn{1}{|c|}{ $C_9$}       &
                \multicolumn{1}{|c|}{$C_{10}$} &
 \multicolumn{1}{|c|}{ $C^{(0)}$ }     \\
        \hline 
        $-0.240$ & $+1.103$ & $+0.011$ & $-0.025$ & $+0.007$ & $-0.030$ &
   $-0.311$ &   $+4.153$ &    $-4.546$    & $+0.381$     \\
        \hline
        \end{tabular}
        \end{center}
\caption{ \it Values of the Wilson coefficients used in the numerical
          calculations corresponding to the central values 
          of the parameters given in Table \protect\ref{parameters}.
For $C_9$ we use the NDR scheme.}
\label{wilson}
\end{table}

\section{Hadron Spectra in \bxsll in the Fermi Motion Model
Including Long-Distance Contribution \label{FMspectra}}

In this section, we study non-perturbative effects
associated with the bound state nature of the $B$ hadron and of the 
$c\bar{c}$
resonances $B \to X_s + (J/\psi,\psi^\prime,...) \to X_s \ell^+ \ell^-$
(the LD contribution) on the hadronic invariant mass and hadron energy 
distributions in the decay \bxsll. Wave-function effects are studied
in the FM model \cite{aliqcd} and for the resonant part we use data.
The FM model parameters are then constrained from the measured 
$J/\psi$-momentum spectrum \cite{CLEOjpsi94}. Sensitivity
of the inclusive spectra in \bxsll on the assumed resonant and
the perturbative $c\bar{c}$-contribution is presented.
The hadronic spectral moments $\langle X_H^n \rangle$, with $X=S,E$ and
$n=1,2$, are calculated numerically in the FM model including the LD-effects.
For the sake of comparison, the corresponding quantities
calculated for the SD-contribution using HQET and the FM model from 
\cite{AH98-2} are also given.
\subsection{Hadron spectra in \bxsll in the Fermi motion model} 

The FM model often invoked in phenomenological studies of $B$ 
decays \cite{aliqcd} has two parameters $p_F$, the Fermi momentum of
the $b$-quark, and the spectator quark 
mass $m_{q}$. Energy-momentum conservation requires the $b$-quark mass
to be a momentum-dependent parameter determined by the constraint:
\begin {eqnarray}
m_b^2(p) = {m_B}^2 + {m_q}^2 -2m_B \sqrt{p^2 + {m_q}^2} 
\quad ; \quad p = |\vec{p}|\, .
\end{eqnarray}
The $b$-quark momentum $p$ is assumed to have a Gaussian 
distribution, denoted by $\phi(p)$, which is determined by $p_F$
 \begin{equation}
\label{lett13}
 \phi(p)= \frac {4}{\sqrt{\pi}{p_F}^3} \exp (\frac {-p^2}{{p_F}^2}) \; ,
\end{equation}
with the normalization
$ \int_0^\infty \, dp \, p^2 \, \phi(p) = 1 $.
 In this
model, the HQET parameters $\bar{\Lambda}$ and $\lambda_1$, representing,
respectively, the binding energy and the kinetic energy of the $b$-quark
inside a $B$ meson,  are calculable in terms of $p_F$ and $m_q$ with 
\begin{eqnarray}
\label{fmtohqet}
\bar{\Lambda} &=& \int_0^\infty dp \, p^2 \phi(p) \sqrt{m_q^2+p^2}, 
\nonumber\\
\lambda_1 &=& -  \int_0^\infty dp \, p^4 \phi(p) = - \frac{3}{2} p_F^2~.
\end{eqnarray}
In addition, for $m_q=0$, one can show that $\bar{\Lambda}=2p_F/\sqrt{\pi}$.
There is, however, no analog of $\lambda_2$ in the FM model.
 For subsequent
use in working out the normalization (decay widths) in the FM model, we
also define an {\it effective} $b$-quark mass by
\begin{equation}
\label{effbmass}
m_b^{{\mbox{eff}}}\equiv(\int_0^\infty  dp \, p^2 \, m_b(p)^5 
\phi(p))^{1/5}~.
\end{equation}
 With the 
quantities $m_b^{\mbox{eff}}$, $\lambda_1$ and $\bar{\Lambda}$ defined 
above, the relation
\begin{equation}
\label{mbfm}
m_B=m_b^{\mbox{eff}}+ \bar{\Lambda} -\lambda_1/(2m_b^{\mbox{eff}})~,
\end{equation}
is found to be satisfied in the FM model to a high accuracy
 (within $0.7 \%$), which is shown in 
Table \ref{tab:FMhqet} for some representative values of the 
HQET parameters and their FM model equivalents. We shall use the 
HQET parameters $\bar{\Lambda}$ and $\lambda_1$ to characterize also the
FM model parameters, with the relations given in eqs.~(\ref{fmtohqet})
and (\ref{effbmass}) and in Table \ref{tab:FMhqet}.  

\begin{table}[h]
        \begin{center}
        \begin{tabular}{|l|l|l|l|}
        \hline
        \multicolumn{1}{|c|}{$p_F,m_q$ (MeV,MeV)}       & 
                \multicolumn{1}{|c|}{$m_b^{{\mbox{eff}}}$ (GeV)} & 
 \multicolumn{1}{|c|}{$\lambda_1$ $(\mbox{GeV}^2)$} &  
\multicolumn{1}{|c|}{$\bar{\Lambda}$ (GeV) } \\
        \hline \hline
        $(450,0)    $     & $4.76 $  & -0.304 & 0.507 \\
        $(252,300)  $    & $4.85 $  & -0.095 & 0.422  \\
        $(310,0)    $      & $4.92 $  & -0.144 & 0.350  \\
\hline
        $(450,150)  $     & $4.73 $  & -0.304 & 0.534 \\
        $(500,150)  $     & $4.68 $  & -0.375 & 0.588  \\
        $(570,150)  $     & $4.60 $  & -0.487 & 0.664  \\
        \hline
        \end{tabular}
        \end{center}
\caption{\it Values of non perturbative parameters
$m_b^{{\mbox{eff}}}$, $\lambda_1$ and $\bar{\Lambda}$  defined in the FM 
model for different sets of the FM model parameters $(p_F,m_q)$.}
\label{tab:FMhqet}
\end{table}

\begin{figure}[htb]
\vskip 0.0truein
\centerline{\epsfysize=3.5in
{\epsffile{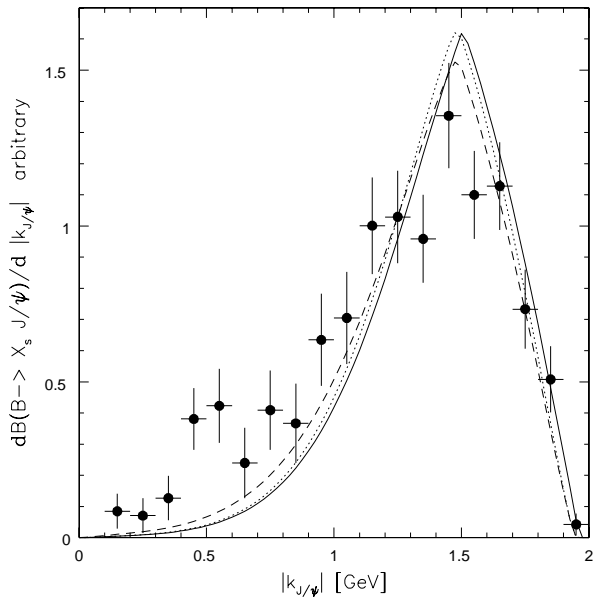}}}
\vskip 0.0truein
\caption[]{ \it Momentum distribution of $J/\psi$ in the decay
$B \to X_s J/\psi$ in the FM model.
The
solid, dotted, dashed curve corresponds to the parameters
$ (\lambda_1, \bar{\Lambda})=(-0.3,0.5),(-0.3,0.53),(-0.38,0.59)$ in
(GeV$^2$, GeV), respectively. The data points are from the CLEO measurements
\cite{CLEOjpsi94}.}
\label{fig:jpsi}
\end{figure}
\begin{figure}[htb]
\vskip 0.0truein
\centerline{\epsfysize=3.5in   
{\epsffile{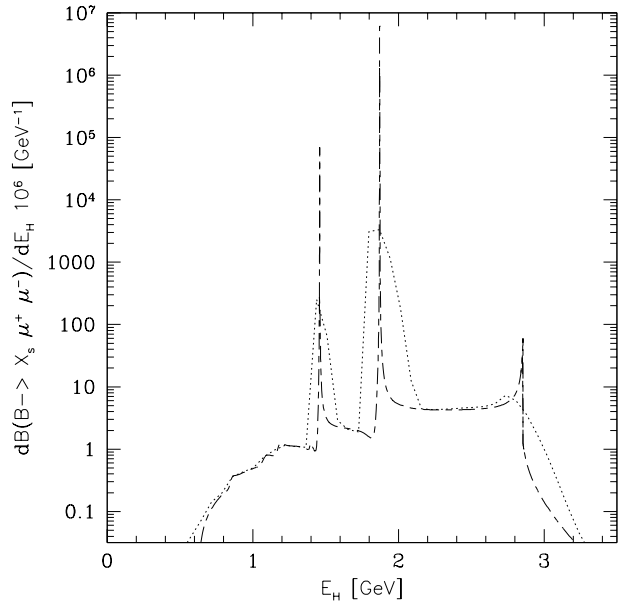}}}
\vskip 0.0truein
\caption[]{ \it Hadron energy spectrum in \bxsll
including the resonant $(J/\psi,\psi^\prime,...)$ and perturbative 
contributions in the Fermi motion model (dotted curve) for $(\lambda_1,
\bar{\Lambda})=(-0.1~\mbox{GeV}^2,0.4~\mbox{GeV})$, and in the parton
model (long-short dashed curve) for $m_b=4.85$ GeV.}
\label{fig:LDEh485}
\end{figure}
\begin{figure}[htb]
\vskip -0.0truein
\centerline{\epsfysize=3.5in  
{\epsffile{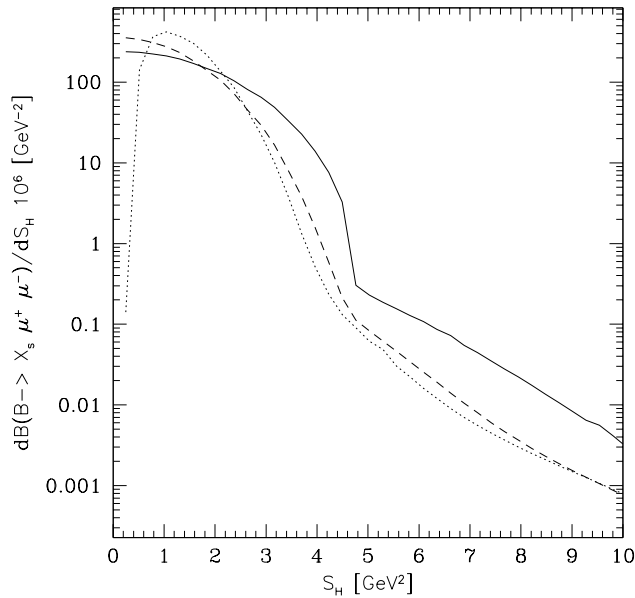}}}
\vskip -0.0truein
\caption[]{ \it Hadronic invariant mass spectrum in \bxsll including the
perturbative and resonant $(J/\psi,\psi^\prime,...)$ contributions in the 
Fermi motion 
model. The solid, dotted and dashed curves correspond to the parameters
$ (\lambda_1, \bar{\Lambda})=(-0.3,0.5),(-0.1,0.4),(-0.15,0.35)$ in
(GeV$^2$, GeV), respectively.}
\label{fig:LDSh} 
\end{figure}
\begin{figure}[t]
     \mbox{ }\hspace{-0.7cm}
     \begin{minipage}[t]{8.2cm}
     \mbox{ }\hfill\hspace{1cm}(a)\hfill\mbox{ }
     \epsfig{file=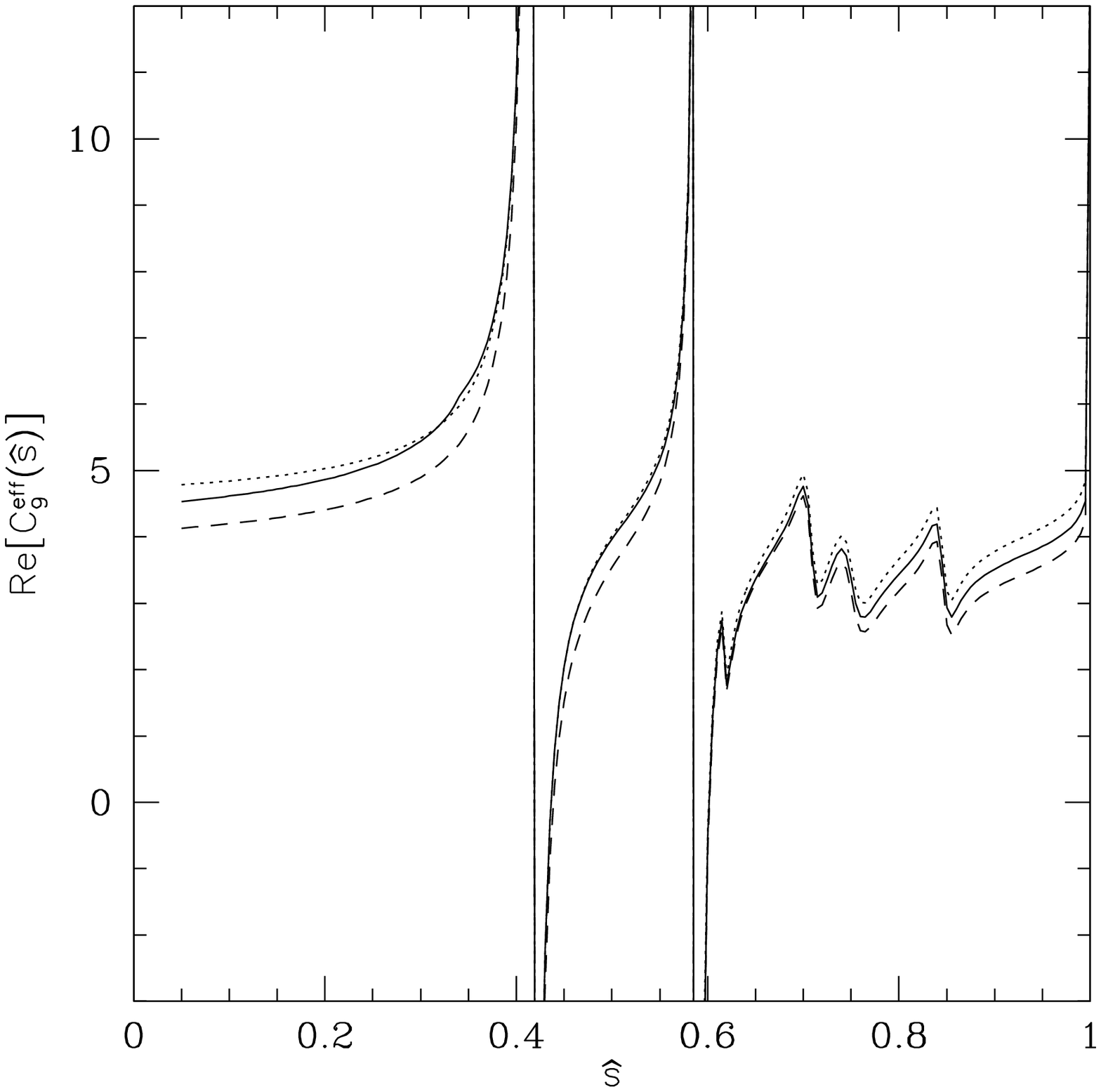,width=8.2cm}
     \end{minipage}
     \hspace{-0.4cm}
     \begin{minipage}[t]{8.2cm}
     \mbox{ }\hfill\hspace{1cm}(b)\hfill\mbox{ }
     \epsfig{file=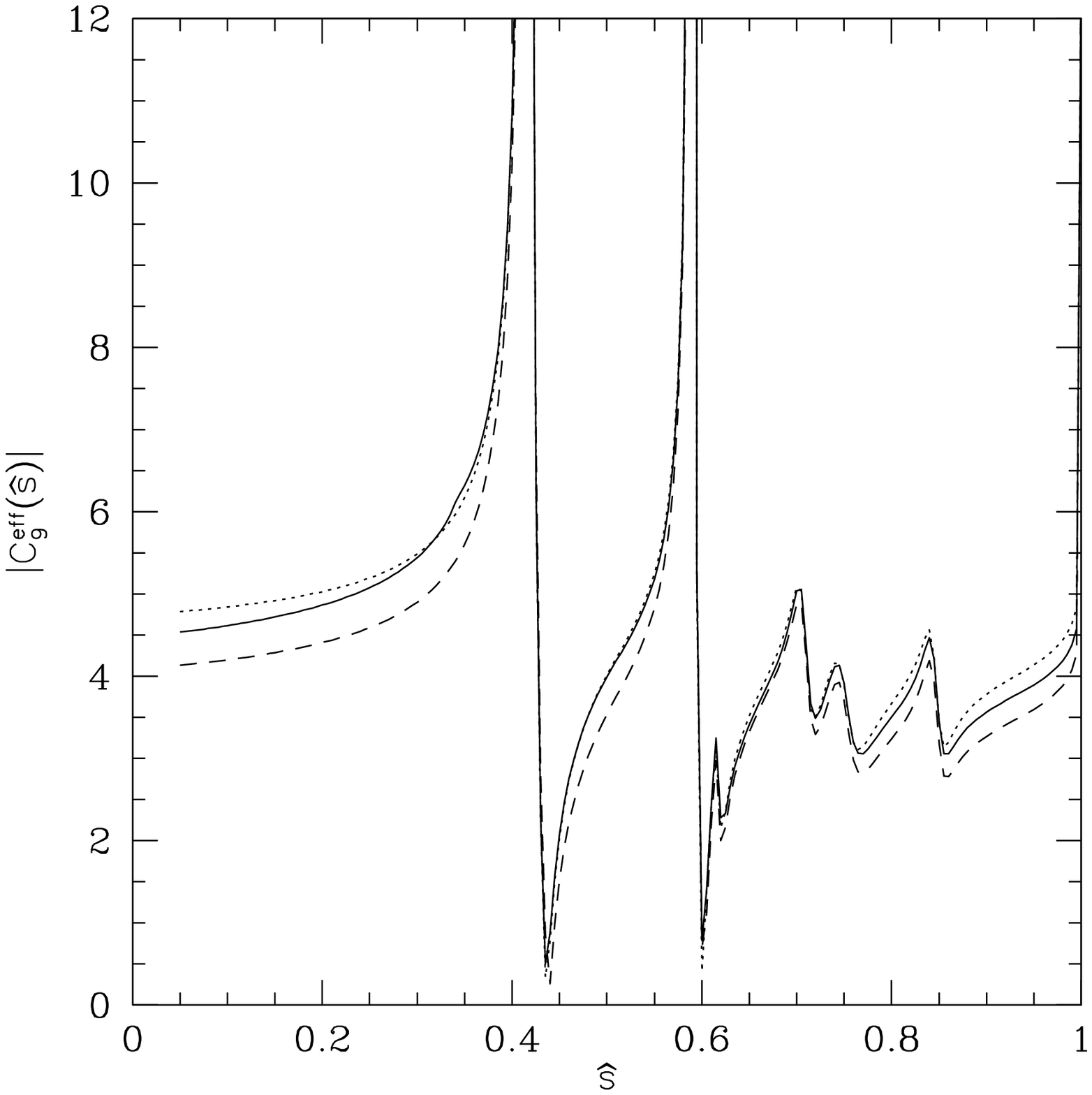,width=8.2cm}
     \end{minipage}  
     \caption{\it 
The real part (a) and the absolute value (b) of $C_9^{{\mbox{\rm eff}}}(\s)$ 
are shown as a function of $\s$, where $C_9^{\mbox{\rm eff}}(\s)=C_9 
\eta(\s) + Y(\s)  +Y_{res}(\s)$. 
The solid curve corresponds to $Y(\s)$ calculated using the
perturbative $c\bar{c}$ contribution $g(\mc,\s)$ given in
eq.~(\ref{gpert}), and the dotted curve corresponds to using 
$\tilde{g}(\mc, \s)$ in eq.~(\ref{eq:gtilde}), with $R_{res}(\s)$ calculated
in both cases using eq.~(\ref{LDeq}). The dashed curve corresponds to
using the Kr\"uger-Sehgal approach \cite{KS96} discussed in the text.}
\label{fig:c9real}
\end{figure}
%
%
%
%
\begin{figure}[t]
     \mbox{ }\hspace{-0.7cm}
     \begin{minipage}[t]{8.2cm}
     \mbox{ }\hfill\hspace{1cm}(a)\hfill\mbox{ }
     \epsfig{file=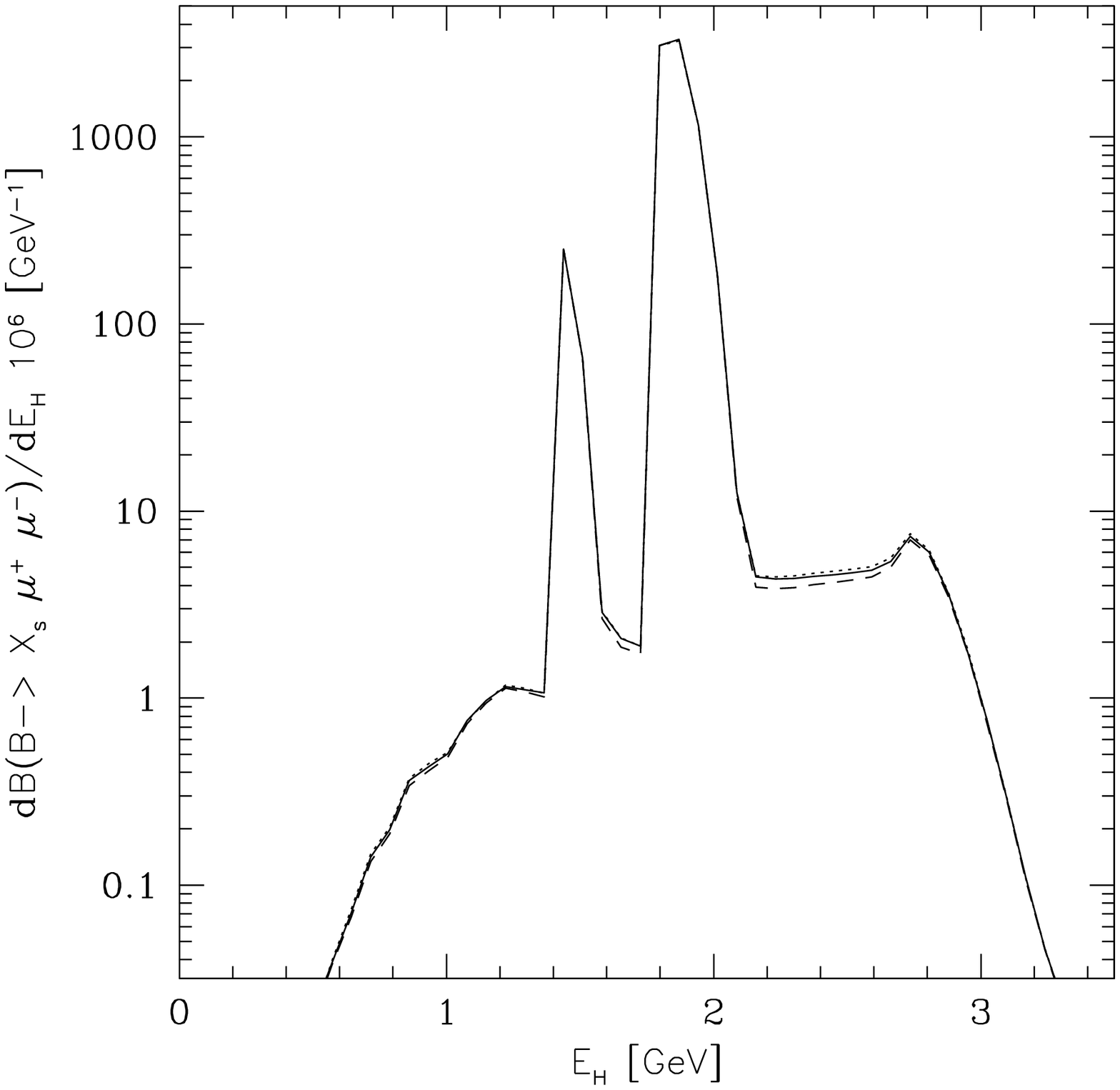,width=8.2cm}
     \end{minipage}
     \hspace{-0.4cm}
     \begin{minipage}[t]{8.2cm}
     \mbox{ }\hfill\hspace{1cm}(b)\hfill\mbox{ }
     \epsfig{file=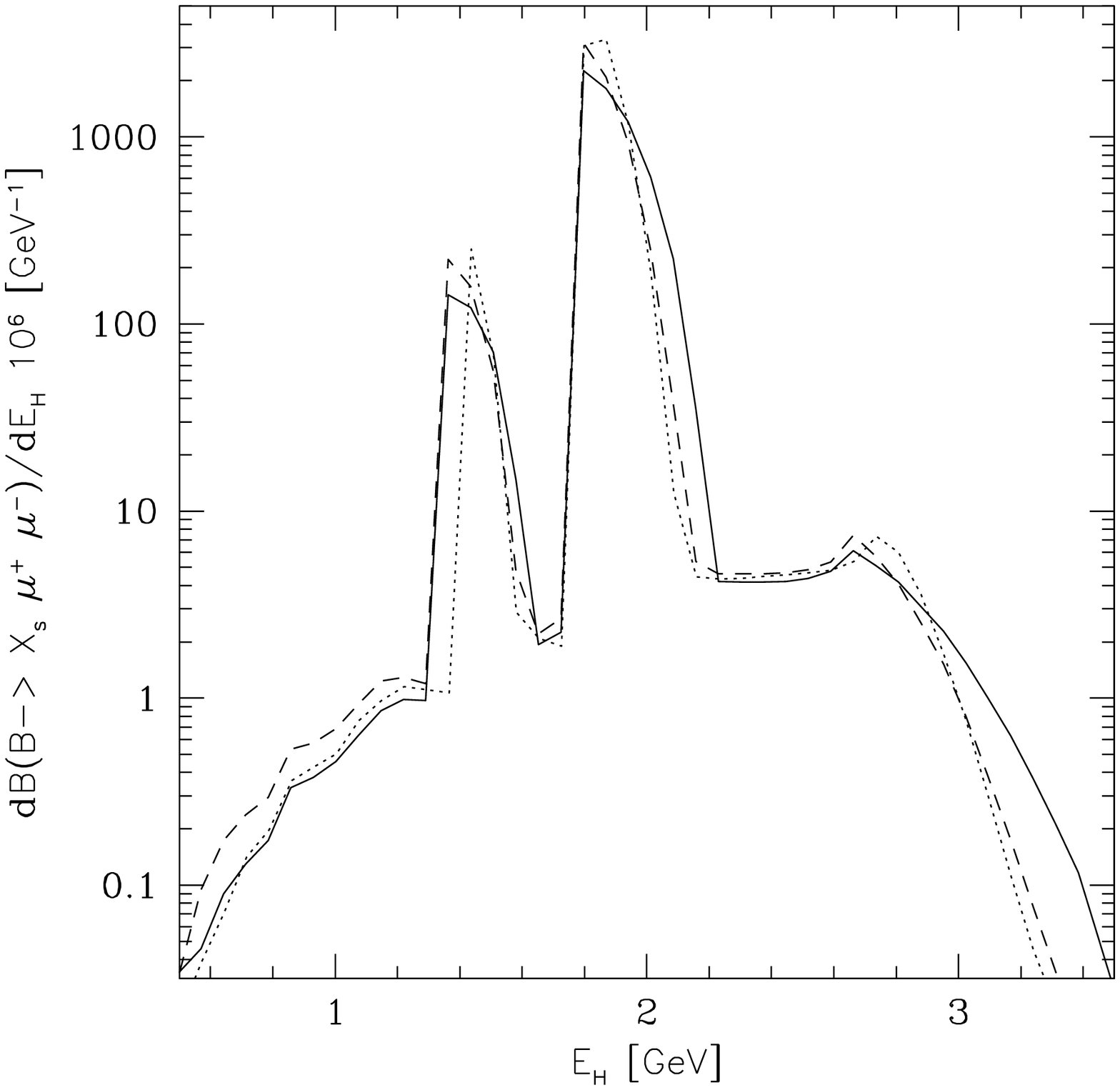,width=8.2cm}
     \end{minipage}  
     \caption{\it 
Hadron energy spectrum in \bxsll including the resonant and perturbative 
contributions in the Fermi motion model.
In (a), the FM model parameters are fixed at $(\lambda_1, 
\bar{\Lambda})=(-0.1~\mbox{GeV}^2,0.4~\mbox{GeV})$.  
The almost overlapping curves differ in the parametrization of 
$C_9^{eff}(\s)$ as depicted in Fig.~(4). The solid curve is obtained using 
eq.~(\ref{gpert}) for $g(\mc,\s)$, the dotted curve is based on 
$\tilde{g}(\mc,\s)$ given in eq.~(\ref{eq:gtilde}), with $R_{res}(\s)$ 
calculated in both cases using eq.~(\ref{LDeq}), and the dashed curve 
corresponds to the Kr\"uger-Sehgal approach \cite{KS96}. 
In (b), the solid, dotted and dashed
curves correspond to the parameters
$ (\lambda_1, \bar{\Lambda})=(-0.3,0.5),(-0.1,0.4),(-0.15,0.35)$ in
(GeV$^2$, GeV), respectively.}
\label{fig:LDEh}
\end{figure}
\subsection{Long-distance resonant contribution in \bxsll}
If not stated otherwise, we shall
follow the procedure adopted in \cite{AHHM97}, in which the 
long-distance (LD) resonance effects in the decay \bxsll (specified below)  
are added with the 
perturbative $c\bar{c}$ contribution expressed through the function
 $g(\hat{m_c},\hat{s})$ in section 2 (see, eq.~(\ref{gpert})). Thus, 
\begin{equation}
\label{simpleadd}
 C_9^{\mbox{eff}}(\s)=C_9 \eta(\s) + Y(\s)  +Y_{res}(\s)~.
 \end{equation}
The function $Y_{res}(\s)$ accounts for the charmonium resonance
 contribution
via $B \to X_s (J/\Psi, \Psi^\prime, \dots) \to X_s \ell^{+} \ell^{-}$ 
for which we take the representation \cite{amm91},
\begin{equation}
        Y_{res}(\s) = \frac{3}{\alpha^2} \kappa \, C^{(0)}
                \sum_{V_i = \psi(1s),..., \psi(6s)}
                \frac{\pi \, \Gamma(V_i \rightarrow \ell^+ \ell^-)\, M_{V_i}}{
                {M_{V_i}}^2 - \s \, {m_b}^2 - i M_{V_i} \Gamma_{V_i}} ,
\label{LDeq}
\end{equation}
where
$C^{(0)} \equiv 3 C_1 + C_2 + 3 C_3 + C_4 + 3 C_5 + C_6$.
We adopt $\kappa = 2.3 $ for the numerical calculations \cite{LW96}.
This is a fair representation of present data in the factorization approach
\cite{NS97}; also the phase of $\kappa$, which is fixed
in eq.~(\ref{LDeq}), is now 
supported by data which finds it
close to its perturbative value \cite{BHP96}.  Note that in this 
approach, the effective coefficient $C_9^{\mbox{eff}}(\hat{s})$ has a
$\hat{s}$-dependence, which is not entirely due to the propagators
in the function $Y_{res}(\hat{s})$ as also  the perturbative
$c\bar{c}$ contribution $g(\hat{m_c}, \hat{s})$ is a function of 
$\hat{s}$. 
In the resonant region, the perturbative part is not noticeable  due to the 
fact that the resonant part in $C_9^{\mbox{eff}}(\hat{s})$ completely 
dominates. However, when 
the $c\bar{c}$ pair is sufficiently off-shell, the
$\hat{s}$-dependence of the function $C_9^{\mbox{eff}}(\hat{s})$
is not (and should not be) entirely determined by the $c\bar{c}$ resonant
contribution. This is the motivation of the representation in 
eq.~({\ref{simpleadd}). We shall later evaluate the
uncertainties in various distributions arising from varying definitions of 
the perturbative contribution $g(\hat{m_c}, \hat{s})$ 
as well as the precise form of the resonating contribution.

\begin{figure}[t]
     \mbox{ }\hspace{-0.7cm}
     \begin{minipage}[t]{8.2cm}
     \mbox{ }\hfill\hspace{1cm}(a)\hfill\mbox{ }
     \epsfig{file=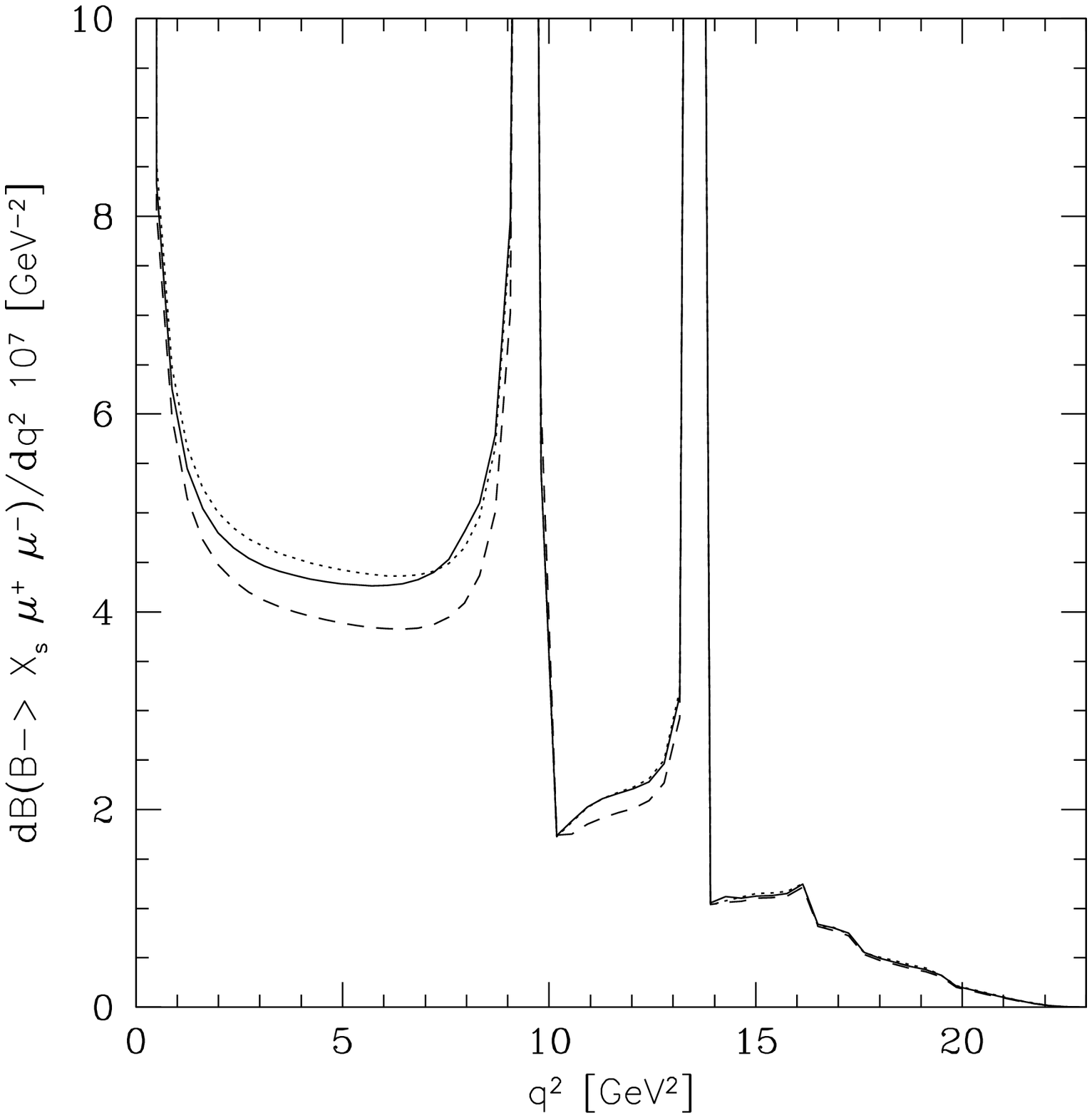,width=8.2cm}
     \end{minipage}
     \hspace{-0.4cm}
     \begin{minipage}[t]{8.2cm}
     \mbox{ }\hfill\hspace{1cm}(b)\hfill\mbox{ }
     \epsfig{file=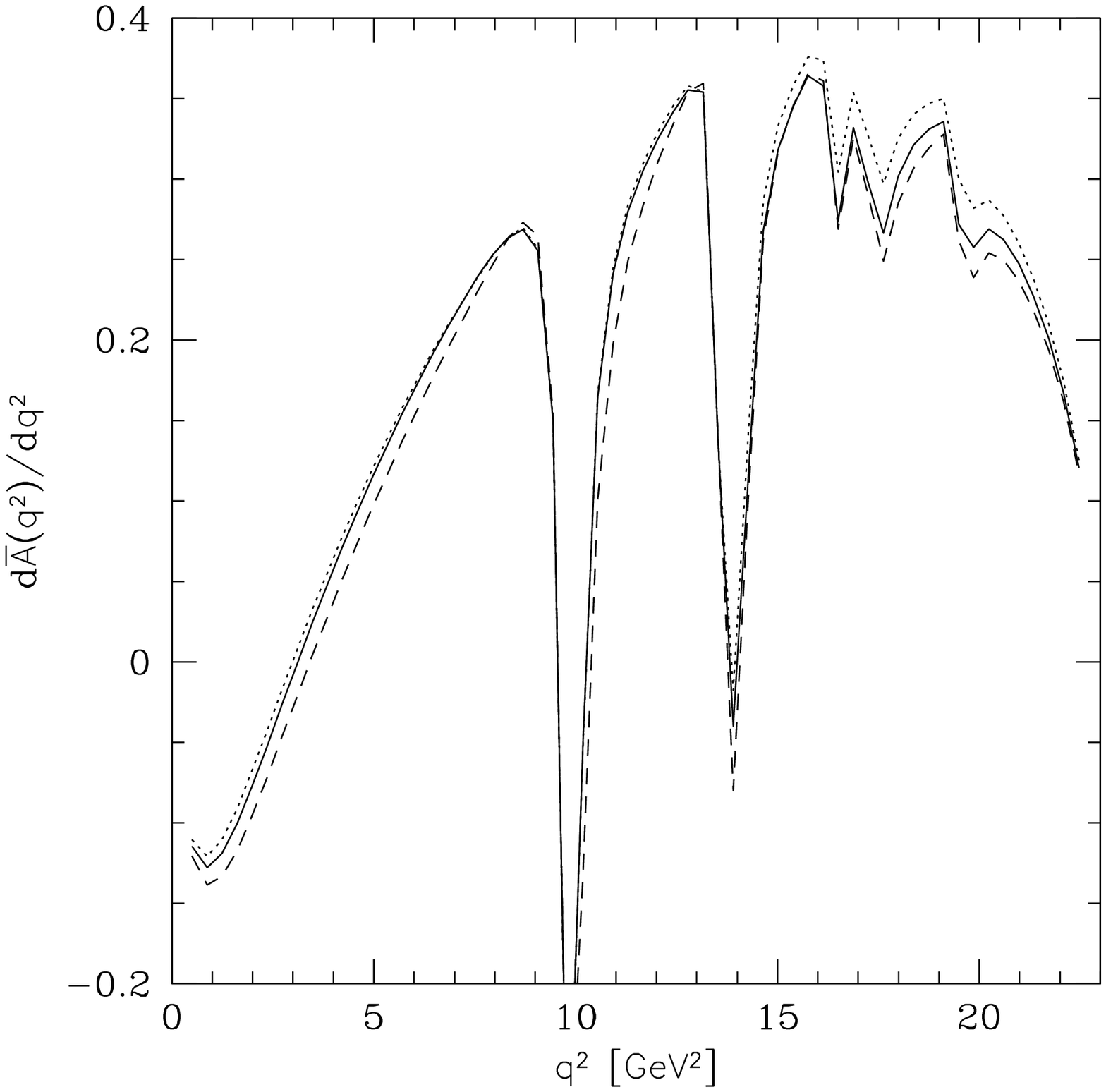,width=8.2cm}
     \end{minipage}  
     \caption{\it 
Dilepton invariant spectrum (a) and the (normalized) 
Forward-Backward asymmetry (b) 
in \bxsll including the resonant and perturbative 
contributions in the Fermi motion model.
The FM model parameters are fixed at $(\lambda_1, 
\bar{\Lambda})=(-0.1~\mbox{GeV}^2,0.4~\mbox{GeV})$.
The solid curve is obtained using  
eq.~(\ref{gpert}) for $g(\mc,\s)$, the dotted curve is based on
$\tilde{g}(\mc,\s)$ given in eq.~(\ref{eq:gtilde}), with $R_{res}(\s)$
calculated in both cases using eq.~(\ref{LDeq}), and the dashed curve
corresponds to the Kr\"uger-Sehgal approach \cite{KS96}.}  
\label{fig:LDshat}
\end{figure}

\subsection{Constraints on the FM model parameters from existing data}
The FM model parameters $p_F$ and $m_q$ (equivalently $\lambda_1$ and
$\bar{\Lambda}$) for the SD-contribution can, in principle, be determined 
from an 
analysis of the energy spectra in the decays $B \to X_u \ell \nu_\ell$
and   $B \to X_s + \gamma$,
as all of them involve the decay of a $b$ quark into (an almost)
massless ($u$ or $s$) quark.
However, the quality of the present data on $B \to X_s + \gamma$ 
does not allow to draw very quantitative conclusions, and hence we
vary the parameters in a reasonable range giving a satisfactory
description of the photon energy spectrum and show the
resulting uncertainties on the hadron spectra in $B \to X_s \ell^+ \ell^-$.
 For estimating the spectra from the LD 
contribution involving the transition $b \to c$, the parameters of the FM 
model can 
be constrained from the lepton energy spectrum in the decay $B \to X_c \ell
\nu_\ell$ and from the shape of the $J/\psi$- and $\psi^\prime$-
momentum distributions in the decays $B \to X_s (J/\psi,\psi^\prime)$.
We review below the presently available analyses of the photon- and 
lepton-energy 
spectra in $B$ decays in the FM model and also present an analysis
of the $J/\psi$-momentum spectrum in $B \to X_s J/\psi$.

\begin{itemize}
\item Analysis of the photon energy spectrum in  $B \to X_s + \gamma$ 
\end{itemize}
The photon energy- and invariant hadronic mass distributions
in $B \to X_s \gamma$ were calculated in the FM model  
using the leading order (in $\alpha_s)$ corrections in ref.~\cite{ag91}.
These spectra were used in the analysis
of the CLEO data on $B \to X_s + \gamma$ \cite{cleobsgamma}, in which the
values $p_F=270 \pm 40$ MeV suggested by the analysis of the CLEO data
on $B \to X \ell \nu_\ell$  were used, together with the effective $b$-quark
mass $m_b^{\mbox{eff}}=4.87 \pm 0.10$ GeV, which gave reasonable fits of the
data. We translate these parameters in terms of  $\lambda_1$
and $\bar{\Lambda}$ using the relations given in eqs.~(\ref{fmtohqet})
and (\ref{mbfm}), yielding
\begin{equation}
\label{cleofmpar}
\lambda_1=-0.11 ^{-0.035}_{+0.030}~\mbox{GeV}^2, ~~~
\bar{\Lambda} = 0.40 \pm 0.1 ~\mbox{GeV}~.
\end{equation} 
The same data was fitted in ref.~\cite{ag95} in the FM model, yielding 
$(p_F,m_q) =(0.45~\mbox{GeV},0~\mbox{GeV})$ as the best-fit
solution, with $(p_F,m_q) =(0.310~\mbox{GeV},0.3~\mbox{GeV})$ differing 
from the best-fit solution by one unit in $\chi^2$.
 The best-fit values translate into
\begin{equation} 
\label{agfmpar}
\lambda_1=-0.3 ~\mbox{GeV}^2~, ~~~\bar{\Lambda}=0.5 ~\mbox{GeV}~.
\end{equation}
Within the indicated errors, the values given in eqs.~(\ref{cleofmpar})
and (\ref{agfmpar}) are compatible. 
\begin{itemize}
\item Analysis of the lepton energy spectrum in  $B \to X \ell \nu_\ell$ 
\end{itemize}
 A fit of the
lepton energy spectrum in the semileptonic decay $B \to X\ell \nu_\ell$
in the context of HQET has been performed in ref.~\cite{gremm}. Using the
CLEO data \cite{CLEObsl96}, the authors of ref.~\cite{gremm} find:
\begin{equation}
\label{hqetgremm}
\lambda_1= -0.19 \pm 0.10 ~\mbox{GeV}^2~,
~\bar{\Lambda}=0.39 \pm 0.11 ~\mbox{GeV} ~.
\end{equation}
Since the FM model and HQET yield very similar lepton energy spectra (apart
from the end-point), one can take the analysis of \cite{gremm} also
holding approximately for the FM model.
\begin{itemize}
\item Analysis of the $J/\psi$-momentum spectrum in  $B \to X_s J/\psi$
\end{itemize}
An analysis
of the $J/\psi$-momentum spectrum in $B \to X_s (J/\psi,\psi^\prime)$
measured by the CLEO collaboration \cite{CLEOjpsi94} 
has been reported in ref.~\cite{PPS97} using the FM model. The authors 
of ref.~\cite{PPS97} addressed both the 
shape and
normalization of the $J/\psi$-data, using the non-relativistic QCD (NRQCD) 
formalism for the inclusive color singlet and color octet charmonium 
production in $B \to X_sJ/\psi$ and the FM model \cite{aliqcd}. The
preferred FM parameters from this analysis are: $(p_F,m_q) 
=(0.57~\mbox{GeV},0.15~\mbox{GeV})$, where $m_q$ only plays a role in 
determining
the position  of the peak but otherwise does not influence the small momentum
tail of the $J/\psi$ momentum distribution. 
 This  yields values of the
parameter $p_F$ which are consistent with the ones obtained in
 ref.~\cite{Hwangetal} $p_F=0.54 ^{+0.16}_{-0.15}$, GeV based on an
analysis of the CLEO data on $B \to X \ell \nu_\ell$ \cite{CLEObsl96}.
The central values of $p_F$ in \cite{Hwangetal} as well as in
\cite{PPS97} correspond to  
$m_b^{\mbox{eff}} \simeq 4.6$ GeV, which is on the lower side of the present 
theoretical estimate of the $m_b$ pole mass, namely $m_b=4.8 \pm 0.2$ GeV
\cite{neubertsachr}. 

  We have redone an analysis of the $J/\psi$-momentum distribution 
in the FM model which
is shown in Fig.~\ref{fig:jpsi}. As also 
discussed in \cite{PPS97}, the low-momentum 
$J/\psi$, in particular in the region $|k_{J/\psi}| \leq 0.6$ GeV, are 
problematic for inclusive decay models, including also the FM model
(see Fig.~1).
The measured $|k_{J/\psi}|$-spectrum appears to have a secondary bump;
an inclusive spectrum  behaving as a Gaussian tail or 
having a power-like behavior $\propto |k_{J/\psi}|^{-\delta}$ in this 
region is hard put to explain this data. There are also suggestions in
literature \cite{BN97} that the spectrum in this region is dominated by
the three-body decay $B \to J/\psi \Lambda \bar{p}$ and hence the bump
reflects the underlying dynamics of this exclusive decay. In view of this,
we have taken out the first six points in the low-$|k_{J/\psi}|$ spectrum
and fitted the FM model parameters in the rest of the 
$|k_{J/\psi}|$-spectrum. The three curves shown correspond to the FM model
parameters $(p_F,m_q) =(0.45~\mbox{GeV}, 0~\mbox{GeV})$ (solid curve),
$(p_F,m_q) =(0.45~\mbox{GeV}, 0.15~\mbox{GeV})$ (dotted curve) and
$(p_F,m_q) =(0.50~\mbox{GeV}, 0.15~\mbox{GeV})$ (dashed curve). They 
all have reasonable $\chi^2$, with $\chi^2/dof=1.6, 1.6$ and 1.1, 
respectively. Excluding also the seventh lowest point, the $\chi^2$
improves marginally, with the resulting $\chi^2$ being $\chi^2/dof=1.4, 
1.4$ and 0.94. Including the sixth point, the fits become slightly
worse. However, they are all acceptable fits. It is interesting that the
best-fit solution of the photon energy spectrum in $B \to X_s + \gamma$,
$(p_F,m_q) =(0.45~\mbox{GeV}, 0~\mbox{GeV})$ \cite{ag95}, is also an 
acceptable fit of the $|k_{J/\psi}|$-data. The corresponding $\lambda_1$, 
$\bar{\Lambda}$ and $m_b$ values from these two analyses are compatible
within $\pm 1 \sigma$ with the HQET-based constraints from the semileptonic 
$B$ decays
\cite{gremm}, quoted above. Thus, the values in eq.~(\ref{agfmpar})
appear to be  a reasonable guess of
the FM model parameters. But, more importantly for the present study, the
phenomenological profile of the LD contribution $B \to X_s 
(J/\psi,\psi^\prime,...) \to X_s \ell^+ \ell^-$ presented here is certainly 
consistent with present data and theoretical constraints. 

\subsection{Effects of the Lorentz boost on the hadron spectra in \bxsll}
We now discuss the $B$-meson
wave function effects in the FM model on the hadron spectra in \bxsll.  
Since the resonances in \bxsll are in the dilepton invariant mass 
variable  $s$
and not in $S_H$, and noting that neither $E_0$ (partonic energy) nor $E_H$
are Lorentz-invariant parameters, it is expected on general grounds that
the effect of the Lorentz boost in the FM model on $E_H$- and
$S_H$-distributions will be more marked
than what was found on the invariant dilepton mass spectrum in
\cite{AHHM97}. We recall that for the dilepton invariant mass, the Lorentz
boost
involved in the FM model (Doppler shift) leaves the spectrum invariant
and there is only a residual effect due to the fact that the $b$-quark
mass in the parton model $m_b$ and the effective $b$-quark mass in the FM
model (called $W(p)$ in \cite{AHHM97} and $m_b^{\mbox{eff}}$ here), are  
different quantities. This
difference $(W(p)-m_b)$ (mass defect) smears the dilepton invariant mass
distribution, but being a subleading effect in $1/m_b$ this effect is
small. Not so in the hadron spectra. In the hadron energy spectrum,
the $c\bar{c}$-resonances, which are narrowly peaked in the
parton model, are broadened by the Lorentz boost of the FM model.
To show this, the hadron energy spectrum in the FM model (dotted
curve) is compared with the spectrum in the parton model (long-short dashed
curve) in Fig.~\ref{fig:LDEh485} for identical values of $m_b$ and
$m_b^{\mbox{eff}}$, taken as $4.85$ GeV.
In terms of the hadronic invariant mass, one finds that
the resonant structure is greatly smeared.
The reason for this behavior is that each $q^2$-bin
contributes to a  range of $E_H$ and $S_H$.
The different $q^2$-regions overlap in  $S_H$ resulting in a smearing of
the resonances over a wide range. This can be seen in   
Fig.~\ref{fig:LDSh} for the hadronic invariant mass. Various curves
illustrate the sensitivity of this spectrum on the FM model parameters.

\subsection{Ambiguities in adding LD and SD contributions in \bxsll}
Since we are simply adding the short-distance (SD) and resonant charmonium 
amplitudes, it can not be ruled out that possibly some double 
counting has crept in in the coefficient $C_9^{\mbox{eff}}(\hat{s})$ 
(once as a continuum $c\bar{c}$ contribution and then
again as $J/\psi,\psi^\prime,...$ resonances).
The question is whether the addition of the $c\bar{c}$-continuum 
and resonating pieces as being done here and in \cite{AHHM97} compromises 
the 
resulting theoretical precision significantly. This can only be studied by
comparing the theoretical scenario in question with other trial 
constructions advocated in the literature.
For example, one could retain in the perturbative function $g(\mc, 
\s)$  just the constant part in
$\s$ by replacing $g(\mc, \s)$ by $\tilde{g}(\mc, \s)$, where
\begin{eqnarray}    
\label{eq:gtilde}   
\tilde{g}(\mc, \s) =
-\frac{8}{9} \ln(\frac{m_b}{\mu})-\frac{8}{9} \ln \mc +\frac{8}{27} \; .
\end{eqnarray}
This function (with $\mu=m_b$) has been proposed in \cite{KS96,LSW97}
as an alternative representation of the $c\bar{c}$ perturbative
contribution and represents the (minimal) short-distance contribution.
To study the difference numerically, we plot both the real 
part ${\rm Re} \, C_9^{\mbox{eff}}(\s)$ and the absolute value  
$|C_9^{\mbox{eff}}(\s)|$ as functions of $\hat{s}$ in
 Fig.~\ref{fig:c9real} by using the complete perturbative 
expression for $g(\mc,\s)$ in eq.~(\ref{eqn:y}) and $\tilde{g}(\mc, \s)$
given in eq.~(\ref{eq:gtilde}). In both cases, the resonant contributions
are included using eq.~(\ref{LDeq}). 
 As a third parametrization of 
$C_9^{\mbox{eff}}(\hat{s})$, we use the  
the approach of Kr\"uger and Sehgal \cite{KS96}, based on  
dispersion relation 
\footnote{The lower limit of the integral
in eq.~(\ref{eq:ks96}) must be below $\hat{m}_{J/\psi}^2$ 
to include the dominant lowest $c \bar{c}$ resonance contributions.
We use $4 \hat{m}_{\pi}^2$ in consultation with F.~Kr\"uger.}
:
\begin{eqnarray}
\label{eq:ks96}
\mbox{Im} ~g(\mc, \s) &=& \frac{\pi}{3} R_{\rm had}^{c\bar{c}}(\s)~,
\nonumber\\
\mbox{Re} ~g(\mc, \s) &=& -\frac{8}{9} \ln \hat{m}_c - \frac{4}{9} 
+\frac{\s}{3} \int_{4\hat{m}_{\pi}^2}^{\infty} \frac{R_{\rm 
had}^{c\bar{c}}(\s)}{\s^\prime (\s^\prime -\s)} d \s^\prime ~.
\end{eqnarray}
The cross-section ratio $R_{\rm had}^{c\bar{c}}(\s)$ in this approach is
expressed as
\begin{equation}
R_{\rm had}^{c\bar{c}}(\s)= R_{cont}^{c\bar{c}}(\s) + R_{\rm 
res}^{c\bar{c}}(\s) ~,
\end{equation}
where $R_{cont}^{c\bar{c}}(\s)$ and $R_{res}^{c\bar{c}}(\s)$ denote the
contribution from the continuum and the narrow resonances, respectively.
For the narrow resonances, the Breit-Wigner form given below 
in eq.~(\ref{eq:Rres}) is used,
whereas for the continuum part a parametrization of the $e^+e^-$ 
annihilation data, taken from \cite{Burkhardt}, is used. 
\begin{eqnarray}
\label{eq:Rres}
R_{res}^{c \bar{c}}(\s) = \kappa \sum_{V_i = \psi(1s),..., \psi(6s)} 
\frac{9 \s}{\alpha^2} 
\frac{m_b^2 Br(V_i \to \ell^+ \ell^-) \Gamma_{total}^{V_i} \Gamma_{had}^{V_i}}
{(\s m_b^2-m_{V_i}^2)^2+m_{V_i}^2 \Gamma_{total}^{V_i 2} } \; .
\end{eqnarray}
Note that the authors of \cite{KS96} use $\kappa=2.35$ in their numerical 
analyses, which is slightly different than the one used in \cite{AHHM97}.
More importantly, this parametrization of $R_{res}(\s)$ has a different
$\s$-dependence than the one given by eq.~(\ref{LDeq}) and it incorporates
the off-shell dependence of the effective $\gamma^* \to V$ vertex,
discussed in \cite{ahmady96}. 

A number of comments on the curves shown in Fig.~\ref{fig:c9real} and the
resulting hadron energy and hadronic invariant mass distributions is in
order.
\begin{itemize}
\item The results for both ${\rm Re} ~C_9^{\mbox{eff}}(\s)$ and the absolute 
value $|C_9^{\mbox{eff}}(\s)|$ plotted as functions of $\hat{s}$ 
show that the functions  corresponding to \cite{KS96} are lower than the
ones used in \cite{AHHM97}, which, in turn, are lower than the ones 
obtained with the prescription given in \cite{LSW97}. 
However, one sees from Figs.~\ref{fig:c9real} (a) 
and \ref{fig:c9real} 
(b) that the differences between these functions are numerically small. 
\item The three
parametrizations discussed  above give almost identical
hadron spectra. The differences between these approaches in the
$E_H$-spectrum are already 
difficult to see, as shown in
Fig.~\ref{fig:LDEh} (a) where we have plotted the $E_H$-spectra in the FM
model for the three parametrizations;
the effect on the hadronic invariant mass is even less noticeable and
hence is not shown. Quantitatively, the maximum difference 
in the hadron energy and the hadronic invariant mass spectra
is $12.1 (4.5) \%$ and $4.1 (2.5) \%$, respectively.
The difference between the approaches in \cite{AHHM97} and 
\cite{KS96} is larger, as given by the first numbers,
than the one between \cite{AHHM97} and \cite{LSW97}, given by the numbers in 
parentheses. However, these maximum differences occur only over a rather 
limited part of the phase space.

\item Other uncertainties on
the  hadronic distribution are much larger, see, for example, 
Fig.~\ref{fig:LDEh} (b) showing the sensitivity of the hadron energy spectra
on the $B$-meson wave-function parameters. In comparison,
the $c\bar{c}$-resonance/continuum related ambiguity is numerically small.  
\end{itemize}

The results presented in Figs.~\ref{fig:LDSh} and  
\ref{fig:LDEh} 
are the principal phenomenological results derived by us in this paper for
the inclusive hadronic invariant mass and hadron energy  spectra in the
decay \bxsll, respectively, and are of direct experimental interest.
 The $S_H$-distribution and moments depend on the FM model 
parameters $p_F$ and $m_q$. In the HQET approach, they depend on the 
parameters
$\bar{\Lambda}$ and $\lambda_1$. Since these parameters are already
constrained by present data, the decay \bxsll can be gainfully used to
determine them more precisely. This was discussed in \cite{AH98-1,AH98-2}.
Here, we have estimated the influence of the LD-resonant contribution.

\subsection{Numerical Estimates of the Hadronic Moments in FM model and
HQET   \label{numerics}}

The similarity of the first two hadronic moments $\langle X_H^n \rangle$,
with $X=E, S$ and $n=1,2$, involving the SD-contribution in the decay 
\bxsll in 
the HQET and FM model descriptions was shown quantitatively in \cite{AH98-2}.
We include these numerical results here for comparison with the corresponding
moments calculated including the LD-contribution in the FM model using
the spectra which we have presented above.
The  moments based on the SD-contribution are defined as:
\begin{eqnarray}
\langle X_H^n\rangle=(\int X_H^n\frac{d{\cal{B}}}{dX_H} dX_H)/{\cal{B}}
\hspace{1cm} {\mbox{for}} \, \, \,\,  X=S,E ~.
\end{eqnarray}
The moments $\langle X_H^n\rangle_{\bar{c} c}$ are defined by taking
into account in addition to the SD-contribution also the contributions 
from
the $c\bar{c}$ resonances. The values of the moments in both the
HQET approach and the FM for $n=1,2$ are shown in
Table~\ref{tab:moments}, with the numbers in the parentheses 
corresponding
to the former. They are based on using the central values of the
parameters given in Table~\ref{parameters}, except for the values of 
$\lambda_1$ and $\bar{\Lambda}$ which are explicitly stated. The 
correspondence between the FM model and HQET parameters is
given in eqs.~(\ref{fmtohqet}). As already stated in \cite{AH98-2}, 
both the HQET and the FM model lead to strikingly similar results 
for the SD-contribution based 
hadronic moments shown in this table. However, the moments
$\langle X_H^n\rangle_{\bar{c} c}$ with $X=S,E$ are significantly 
lower
than their SD-counterparts $\langle X_H^n\rangle$ calculated for the
same values of the FM model parameters. This shows
that the $c\bar{c}$ resonances are important also in
moments. The hadronic invariant mass spectra in \bxsll for both the SD
and inclusive (LD+SD) contributions are expected to be dominated by 
multi-body
states, with $\langle S_H\rangle \simeq (1.5 - 2.1) \, \mbox{GeV}^2$ and
$\langle S_H\rangle_{\bar{c} c} \simeq (1.2 - 1.5) \, \mbox{GeV}^2$.

\begin{table}[h]
        \begin{center}   
        \begin{tabular}{|c|l|l|l|l|}
        \hline
        \multicolumn{1}{|c|}{{\mbox{}}}      &
                \multicolumn{1}{|c|}{$\langle S_H\rangle$  } &
\multicolumn{1}{|c|}{$\langle S_H\rangle_{\bar{c} c}$ } &
                \multicolumn{1}{|c|}{$\langle S_H^2\rangle$  } &
\multicolumn{1}{|c|}{$\langle S_H^2\rangle_{\bar{c} c}$ } \\
 \hline
\multicolumn{1}{|c|}{($\lambda_1,\bar{\Lambda})$ in (GeV$^2$, GeV)} &
\multicolumn{2}{|c|}{$({\mbox{GeV}}^2)$ } &
\multicolumn{2}{|c|}{$({\mbox{GeV}}^4)$ } \\
        \hline
    $(-0.3,0.5)$  & 2.03 (2.09)&1.51 &6.43 (6.93)&3.10   \\
    $(-0.1,0.4)$  & 1.75 (1.80)&1.36 &4.04 (4.38)&2.17   \\
    $(-0.14,0.35)$  & 1.54 (1.49)&1.19 &3.65 (3.64)&1.92   \\
        \hline
\hline
        \multicolumn{1}{|c|}{{\mbox{}}} &
               \multicolumn{1}{|c|}{$\langle E_H\rangle$  } &
\multicolumn{1}{|c|}{$\langle E_H\rangle_{\bar{c} c} $} &
                \multicolumn{1}{|c|}{$\langle E_H^2\rangle$  } &
\multicolumn{1}{|c|}{$\langle E_H^2\rangle_{\bar{c} c}$ } \\
 \hline
\multicolumn{1}{|c|}{$(\lambda_1,\bar{\Lambda})$ in (GeV$^2$, GeV)} & 
\multicolumn{2}{|c|}{$({\mbox{GeV)}} $ } &
\multicolumn{2}{|c|}{$({\mbox{GeV}}^2)$ }
\\
        \hline
$(-0.3,0.5)$   &2.23 (2.28)&1.87 &5.27 (5.46)& 3.52   \\
$(-0.1,0.4)$  &2.21 (2.22)&1.85 &5.19 (5.23)& 3.43   \\
$(-0.14,0.35)$  &2.15 (2.18)&1.84 &4.94 (5.04)& 3.39   \\
        \hline
        \end{tabular}
        \end{center}
\caption{\it Hadronic spectral moments for $B \to X_s \mu^{+} \mu^{-}$
in the Fermi motion model (HQET) for the indicated
values of the parameters $(\lambda_1,\bar{\Lambda})$.
 }
\label{tab:moments}
\end{table}
\section{Branching Ratios and Hadron Spectra in \bxsll with Cuts on 
Invariant Masses}
 In experimental searches for the decay \bxsll, the 
short-distance contribution (electroweak penguins
and boxes) is expected to be visible away from the resonances.
So, cuts on the invariant dilepton mass are imposed to stay away from  
the dilpeton mass range where the charmonium resonances $J/\psi$ and
$\psi^{\prime}$ are dominant. 
 For example, the cuts imposed in the recent CLEO analysis
\cite{cleobsll97} given below are typical: 
\begin{eqnarray}
{\mbox{cut A}}&:& 
q^2 \leq  (m_{J/\psi}-0.1 \, {\mbox{GeV}})^2 = 8.98 \, {\mbox{GeV}}^2 \, ,
\nonumber \\
{\mbox{cut B }}&:& 
q^2 \leq  (m_{J/\psi}-0.3 \, {\mbox{GeV}})^2 = 7.82 \, {\mbox{GeV}}^2 \, ,
\nonumber \\
{\mbox{cut C}}&:& 
q^2 \geq  (m_{\psi^{\prime}}+0.1 \, {\mbox{GeV}})^2 = 14.33 \, {\mbox{GeV}}^2
\, . \label{eq:cuts}
\end{eqnarray}
The cuts $A$ and $B$ have been chosen to take into account the
QED radiative corrections as these effects are different in the
$e^+ e^-$ and $\mu^+ \mu^{-}$ modes. In the following, we compare the
hadron spectra with and without the resonances after imposing these
experimental cuts.
For the low-$q^2$ cut for muons (cut $A$), the hadron energy spectra and the
hadronic invariant mass spectra are shown  
in Fig.~\ref{fig:EhLO} (a), (b) and Fig.~\ref{fig:ShLO} (a), (b), respectively.
The results for the low-$q^2$ cut for electrons (cut $B$), are shown in
Fig.~\ref{fig:EhLO} (c), (d) and Fig.~\ref{fig:ShLO} (c), (d), respectively. 
Finally, the hadronic spectra for the high-$q^2$ cut (cut C) for 
$e^{+} e^{-}$ and $\mu^{+} \mu^{-}$ can be seen in 
Fig.~\ref{fig:EhLO} (e), (f) for the hadronic energy and in 
Fig.~\ref{fig:ShLO} (e), (f)
for the hadronic invariant mass.
We see that the above  
cuts in $q^2$ greatly reduce the resonance contributions.
Hence, the resulting distributions essentially test the non-resonant 
$c\bar{c}$ and short-distance contributions. These figures will be used
later to quantify the model dependence of the integrated branching ratios
in \bxsll.

 As mentioned in \cite{cleobsll97}, the dominant $B\bar{B}$ background 
to the decay \bxsll comes from two semileptonic decays of $B$ or $D$
mesons, which produce the lepton pair with two undetected neutrinos.
 To suppress this $B\bar{B}$ background,
it is required that the invariant mass of the final hadronic state is 
less than $t=1.8 \, {\mbox{GeV}}$, which approximately equals $m_D$. 
We define the survival probability of the \bxsll signal 
after the hadronic invariant mass cut:
\begin{eqnarray}
S(t)\equiv (\int_{m_{X}^2}^{t^2} \frac{d{\cal{B}}}{dS_H} dS_H)/{\cal{B}}~,
\label{eq:eff}
\end{eqnarray}
and present $S(t=1.8 ~\mbox{GeV})$ as the
fraction of the branching ratio for \bxsll surviving these cuts  
in Table~\ref{SHoutcome}. To estimate the model dependence
of this probability, we vary the FM 
model parameters. Concentrating on the SD piece, we note that the effect
of this cut alone is that between $83\%$ to $92\%$ of the signal for
$B \to X_s \mu^+ \mu^-$ and between $79\%$ to $90\%$ of the signal
in $B \to X_s e^+ e^-$ survives, depending on the FM model parameters.
The corresponding numbers for the inclusive spectrum including the
SD and LD contribution is $96\%$ to $99.7\%$ for both the dimuon and 
dielectron case. This shows that while this cut removes a good fraction of
the $B\bar{B}$ background, it allows a very large fraction of the \bxsll
signal to survive. However, this cut does not discriminate between the
SD and (SD+LD) contributions, for which the cuts $A$ - $C$ are effective.
The numbers for the survival probability $S(t=1.8 ~\mbox{GeV})$ reflect 
that the hadronic 
invariant mass distribution of the LD-contribution is more steep than
the one from the SD contribution.

  With the additional cut A (B) imposed on the dimuon (dielectron) 
invariant 
mass, between $57\%$ to $65\%$ ($57\%$ to $68\%$) of the \bxsll signal 
survives the additional cut
on the hadronic invariant mass for the SD contribution. However, as
expected, the cuts A and B result in drastic reduction of the inclusive
branching ratio for the decay \bxsll, as they effectively remove the 
dominant $c\bar{c}$-resonant part. In this case only $0.8\%$ to $0.9\%$
($1.0\%$ to $1.2\%$) of the inclusive signal survives for the cut A (B).
 The theoretical branching ratios for both the dielectron 
and dimuon
cases, calculated using the central values in Table ~\ref{parameters}
and the indicated values of $\lambda_1$ and $\bar{\Lambda}$ are 
also given in Table~\ref{SHoutcome}. As estimated in \cite{AH98-2}, the 
uncertainty on the branching ratios resulting from the errors on the
parameters in Table~\ref{parameters} is about $\pm 23\%$ (for the dielectron 
mode) and
$\pm 16 \%$ (for the dimuon case). The wave-function-related uncertainty in
the branching ratios is smaller, as can be seen in Table ~\ref{SHoutcome}.
This gives a fair estimate of the theoretical uncertainties on   
the partially integrated branching ratios from the $B$-meson wave
function and $c\bar{c}$ resonances.
 With the help of the theoretical branching ratio and the survival
probability $S(t=1.8 \, \mbox{GeV})$, calculated for three sets of the FM 
parameters, the cross section can be calculated for all six 
cases:\\
 (i) no cut on the dimuon invariant mass [(SD) and (SD + LD)],
(ii) no cut on the dielectron invariant mass [(SD) and (SD + LD)],
(iii) cut A on the dimuon invariant mass [(SD) and (SD + LD)],
(iv) cut B on the dielectron invariant mass [(SD) and (SD + LD)],
(v)  cut C on the dimuon invariant mass [(SD) and (SD + LD)],
(vi)  cut C on the dielectron invariant mass [(SD) and (SD + LD)].
This table shows that with $10^7$ $B\bar{B}$ events, ${\cal O}(70)$
dimuon and ${\cal O}(100)$ dielectron signal events from 
\bxsll should survive the CLEO cuts A and B, respectively, with $m(X_s) 
<1.8$ GeV. With the cut C, one expects an order of magnitude less events,
making this region interesting for the LHC experiments which will have
much higher $B\bar{B}$ statistics. Given enough data, one can compare
the experimental distributions in \bxsll directly with the ones presented 
here. The phenomenological success of the FM model in describing 
the energy spectra in
$B$ decays and its close proximity to HQET make us confident that
the hadron spectra in \bxsll presented here should be good descriptions
of the data.    

\subsection{Hadronic Spectral Moments with Cuts in the FM}

We have calculated the first two moments of the hadronic invariant mass in 
the FM model by imposing a cut $S_H < t^2$ with $t=1.8 \, \mbox{GeV}$ 
and an optional cut on $q^2$.
\begin{eqnarray}
\label{eq:SHcut}
\langle S_H^n\rangle=
(\int_{m_{X}^2}^{t^2} S_H^n\frac{d^2{\cal{B}}_{cut X}}{dS_Hdq^2} dS_Hdq^2)
/(\int_{m_{X}^2}^{t^2} \frac{d^2{\cal{B}}_{cut X}}{dS_Hdq^2} dS_Hdq^2)
\hspace{1cm} {\mbox{for}} \, \, \,\,  n=1,2 ~.
\end{eqnarray}  
Here the subscript $cut X$ indicates whether we evaluated 
$\langle S_H\rangle$ and $\langle S_H^2 \rangle$ with the cuts 
on the invariant dilepton mass as defined in 
eq.~(\ref{eq:cuts}), or without any cut on the dilepton mass. 
The results are collected in Table \ref{SHcutmoments}.
The moments given in Table \ref{SHcutmoments} can be
compared directly with the data to extract the FM model parameters.
The entries in this table  give a fairly
good idea of what the effects of the experimental cuts on the corresponding
moments in HQET will be, as the FM and HQET yield very similar moments
for equivalent values of the parameters. The 
functional dependence of 
the hadronic moments on the HQET parameters taking into account the 
experimental cuts still remains to be worked out. 

In the last row of Table \ref{SHcutmoments} the value in percentage
refers to the maximum uncertainty in $\langle S_H^n \rangle$, 
with $n=1,2$, resulting 
from  different approaches to include
the resonant $c \bar{c}$ effects. We have calculated $\langle S_H\rangle$ 
and $\langle S_H^2 \rangle$ with all the cuts mentioned above for the three 
approaches for fixed $(\lambda_1,\bar{\Lambda})=(-0.1,0.4)$ in 
$\mbox{GeV}^2,\mbox{GeV}$. Thus, for example, $\langle S_H \rangle = (1.77
\pm 0.90 \%)~\mbox{GeV}^2$ for Cut A. This uncertainty is much below the
one due to the variations in the parameters $\lambda_1$ and $\bar{\Lambda}$.
Hence, the measurement of the hadronic moments can be used to determine
these parameters. 
  
\begin{table}[h]
        \begin{center}
        \begin{tabular}{|l|l|l|l|l|l|l|l|l|}
        \hline
        \multicolumn{1}{|c|}{FM parameters}       & 
\multicolumn{1}{|c|}{ $ {\cal{B}}\cdot 10^{-6}$} & 
\multicolumn{1}{|c|}{${\cal{B}}\cdot 10^{-6}$} &  
\multicolumn{1}{|c|}{No $s$-cut} & 
\multicolumn{1}{|c|}{No $s$-cut } &  
\multicolumn{1}{|c|}{cut A}  &
\multicolumn{1}{|c|}{cut B}  &
\multicolumn{1}{|c|}{cut C}  &
\multicolumn{1}{|c|}{cut C}  \\
 \multicolumn{1}{|c|}{($\lambda_1,\bar{\Lambda})$ in (GeV$^2$, 
GeV)}           & 
 \multicolumn{1}{|c|}{$\mu^{+} \mu^{-}$}   & 
 \multicolumn{1}{|c|}{$e^+ e^-$}           &  
 \multicolumn{1}{|c|}{$\mu^{+} \mu^{-}$}   & 
 \multicolumn{1}{|c|}{$e^+ e^-$}           &  
 \multicolumn{1}{|c|}{$\mu^{+} \mu^{-}$}   &
 \multicolumn{1}{|c|}{$e^+ e^-$}           &
 \multicolumn{1}{|c|}{$\mu^{+} \mu^{-}$}   &
 \multicolumn{1}{|c|}{$e^+ e^-$}           \\
        \hline \hline
   $(-0.3,0.5) ~[SD] $  &5.8 &8.6 & $83 \% $  & 79 \%&$57 \% $  & $57 \% $ 
&$6.4\% $&$4.5\%$\\
   $(-0.1,0.4) ~[SD]$ &5.7 &8.4 & $93 \% $  & 91 \%&$63 \% $  & $68 \% $
&$8.3\%$&$5.8\%$\\
   $(-0.14,0.35) ~[SD] $  &5.6 &8.3 & $92 \% $  & 90 \%&$65 \% $  & $67 \% $ 
&$7.9\%$&$5.5\%$\\
   $(-0.3,0.5) ~[SD+LD] $ &562.5&563.9 & $96 \% $  & 96 \%&$0.8 \% $ & $1.0 
\% $ &$0.06\%$&$0.06\%$\\
   $(-0.1,0.4) ~[SD+LD]$&564.0&565.6 & $99.7 \% $&99.7\%&$0.8 \% $ & $1.1 
\% $ &$0.08\%$&$0.08\%$\\
   $(-0.14,0.35)~[SD+LD] $ &566.5&568.2 & $99 \% $  & 99 \%&$0.9 \% $ & $1.2 
\% $ &$0.08\%$&$0.08\%$\\
        \hline
        \end{tabular}
        \end{center}
\caption{\it Branching ratios and survival probabilities for \bxsll, 
$\ell=\mu,e$ for different FM model 
parameters evaluated from the SD and $[SD +LD]$ contributions. The 
branching ratios without experimental cuts are given in the second and third
columns. The values given in percentage in the fourth to eleventh columns
represent the  survival probability $S(t=1.8 {\mbox{ {\rm {GeV}}}})$  
defined in eq.~(\ref{eq:eff}) without any cut on the dilepton invariant mass
and for three different cuts as defined in eq.~(\ref{eq:cuts}).}
\label{SHoutcome}
\end{table}
\begin{table}[h]
        \begin{center}
        \begin{tabular}{|l|l|l|l|l|l|l|l|l|l|l|}
        \hline
        \multicolumn{1}{|c|}{FM}       & 
\multicolumn{2}{|c|}{No $s$-cut} & 
\multicolumn{2}{|c|}{No $s$-cut } &  
\multicolumn{2}{|c|}{cut A}  &
\multicolumn{2}{|c|}{cut B}  &
\multicolumn{2}{|c|}{cut C}  \\
 \multicolumn{1}{|c|}{parameters}           & 
 \multicolumn{2}{|c|}{$\mu^{+} \mu^{-}$}   & 
 \multicolumn{2}{|c|}{$e^+ e^-$}           &  
 \multicolumn{2}{|c|}{$\mu^{+} \mu^{-}$}   & 
 \multicolumn{2}{|c|}{$e^+ e^-$}           &  
 \multicolumn{2}{|c|}{$\ell^{+} \ell^{-}$}   \\
 \multicolumn{1}{|c|}{($\lambda_1,\bar{\Lambda})$}           & 
 \multicolumn{1}{|c|}{$\langle S_H \rangle$ } &
\multicolumn{1}{|c|}{$\langle S_H^2\rangle$} &
 \multicolumn{1}{|c|}{$\langle S_H\rangle$  } &
\multicolumn{1}{|c|}{$\langle S_H^2\rangle$ } &
 \multicolumn{1}{|c|}{$\langle S_H\rangle$  } &
\multicolumn{1}{|c|}{$\langle S_H^2\rangle$ } &
 \multicolumn{1}{|c|}{$\langle S_H\rangle$  } &
\multicolumn{1}{|c|}{$\langle S_H^2\rangle$ } &
 \multicolumn{1}{|c|}{$\langle S_H\rangle$  } &
\multicolumn{1}{|c|}{$\langle S_H^2\rangle$ } \\
 \multicolumn{1}{|c|}{GeV$^2$, GeV}           & 
\multicolumn{1}{|c|}{${\mbox{GeV}}^2$ } &
\multicolumn{1}{|c|}{${\mbox{GeV}}^4$ } &
\multicolumn{1}{|c|}{${\mbox{GeV}}^2$ } &
\multicolumn{1}{|c|}{${\mbox{GeV}}^4$ } &
\multicolumn{1}{|c|}{${\mbox{GeV}}^2$ } &
\multicolumn{1}{|c|}{${\mbox{GeV}}^4$ } &
\multicolumn{1}{|c|}{${\mbox{GeV}}^2$ } &
\multicolumn{1}{|c|}{${\mbox{GeV}}^4$ } &
\multicolumn{1}{|c|}{${\mbox{GeV}}^2$ } &
\multicolumn{1}{|c|}{${\mbox{GeV}}^4$ } \\
        \hline \hline
$(-0.3,0.5)  $ &1.47&2.87&1.52&3.05&1.62&3.37&1.66&3.48&0.74&0.69\\
$(-0.1,0.4)  $ &1.57&2.98&1.69&3.37&1.80&3.71&1.88&3.99&0.74&0.63 \\
$(-0.14,0.35)$ &1.31&2.34&1.38&2.55&1.47&2.83&1.52&2.97&0.66&0.54\\
        \hline
$(-0.3,0.5)_{tot}  $ &1.41&2.61&1.41&2.62&1.61&3.32&1.66&3.47&0.74&0.68\\
$(-0.1,0.4)_{tot}  $ &1.35&2.14&1.36&2.15&1.77&3.60&1.87&3.94&0.74&0.62 \\
$(-0.14,0.35)_{tot}$ &1.17&1.84&1.18&1.85&1.45&2.76&1.51&2.95&0.66&0.54\\
        \hline
$\triangle$        &$0.15 \%$&$0.19 \%$&$0.22 \%$&$0.42 \%$&$0.90 \%$&$1.56 \%$&$0.32 \%$&$0.58 \%$&$0.01 \%$&$ 0.32 \%$\\
\hline
        \end{tabular}
        \end{center}
\caption{\it Spectral moments $\langle S_H\rangle$ and $\langle 
S_H^2\rangle$ for \bxsll, $\ell=\mu,e$ for different FM model 
parameters and a hadronic invariant mass cut  $S_H <3.24 \, \mbox{GeV}^2$ 
are given in the second to fifth columns. 
The values in the sixth to eleventh columns have additional cuts on the 
dilepton invariant mass spectrum as defined in eq.~(\ref{eq:cuts}).
The $S_H$-moments with cuts are defined in eq.~(\ref{eq:SHcut}). Entries in 
the first three rows are calculated using the SD-contribution alone. The 
subscript 
$tot=SD+LD$ denotes that both the short and the long distance contribution
are included in these moments. The values of $\triangle$ given in the 
last row represent the maximum uncertainty on the spectral moments (in $\%$)
resulting from the three approaches to take into account the
continuum/$c\bar{c}$-resonant contributions discussed in the text.}
\label{SHcutmoments}
\end{table}

%
%
%
%
%
%
\begin{figure}[t]
\mbox{}\vspace{-2cm}\\
\begin{center}
     \mbox{ }\hspace{-0.7cm}
     \begin{minipage}[t]{7.0cm}
     \mbox{ }\hfill\hspace{1cm}(a)\hfill\mbox{ }
     \epsfig{file=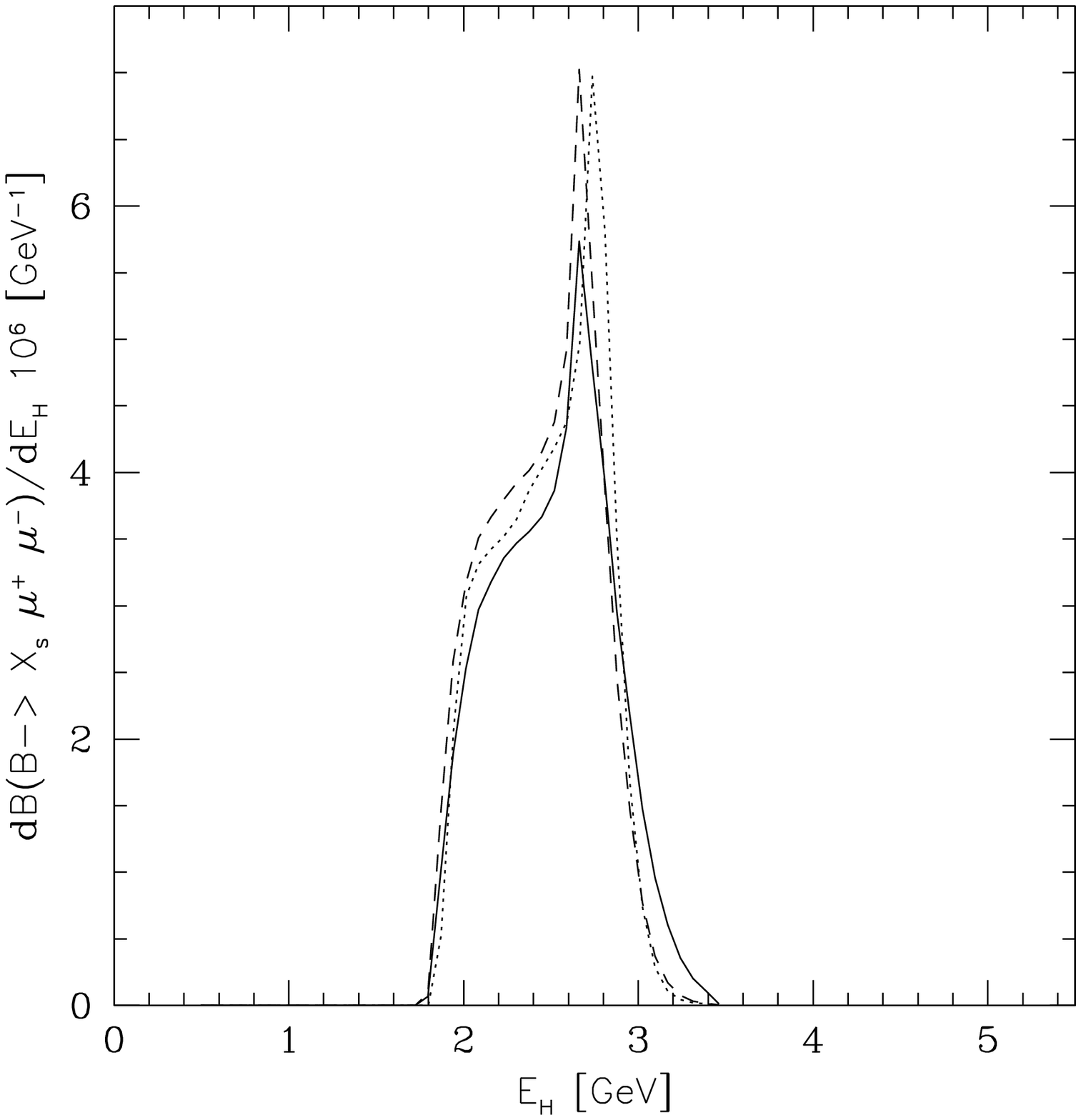,width=7.0cm}
     \end{minipage}
     \hspace{-0.4cm}
     \begin{minipage}[t]{7.0cm}
     \mbox{ }\hfill\hspace{1cm}(b)\hfill\mbox{ }
     \epsfig{file=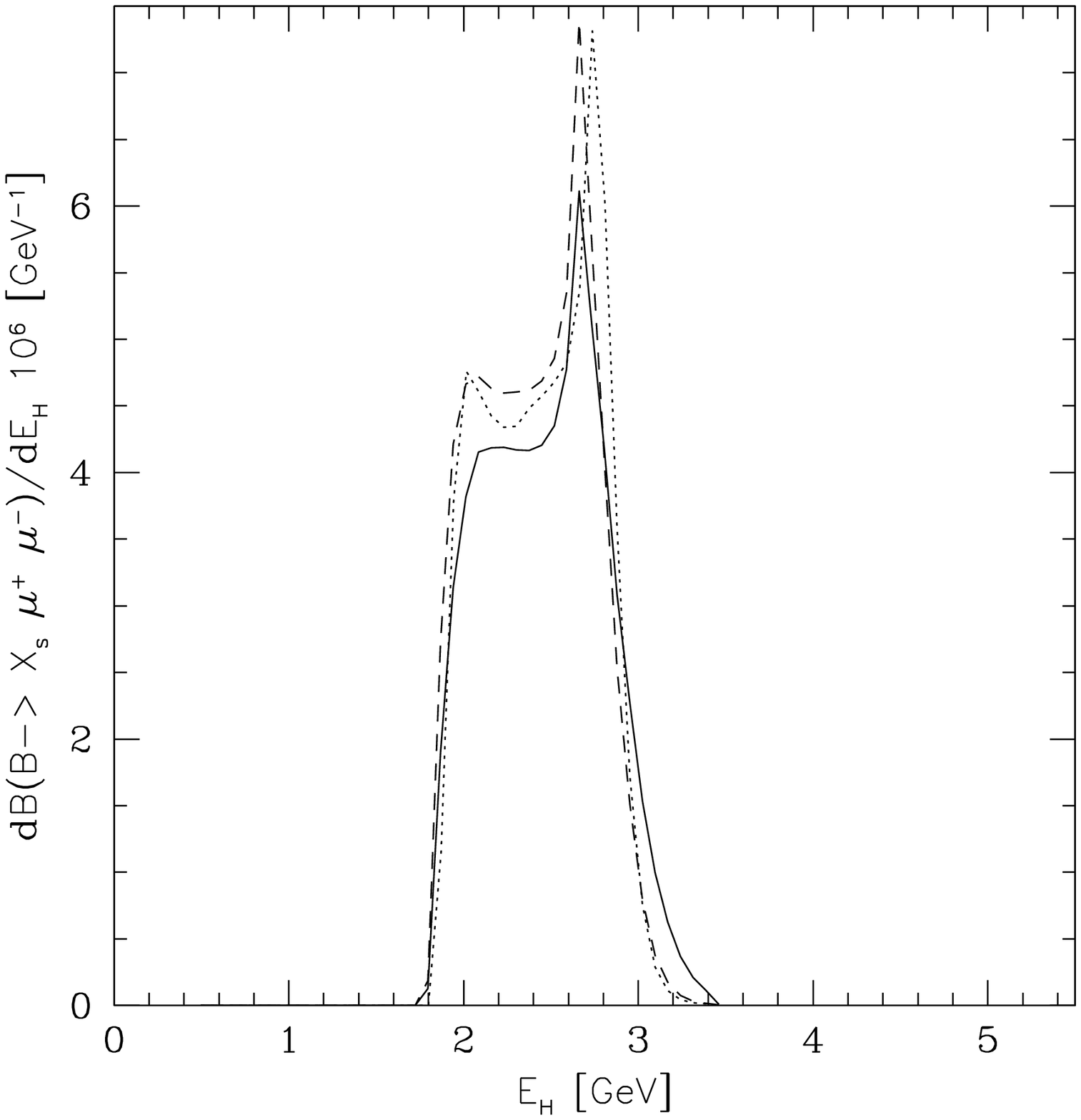,width=7.0cm}
     \end{minipage} \\ 
 \mbox{ }\hspace{-0.7cm}
     \begin{minipage}[t]{7.0cm}
     \mbox{ }\hfill\hspace{1cm}(c)\hfill\mbox{ }
     \epsfig{file=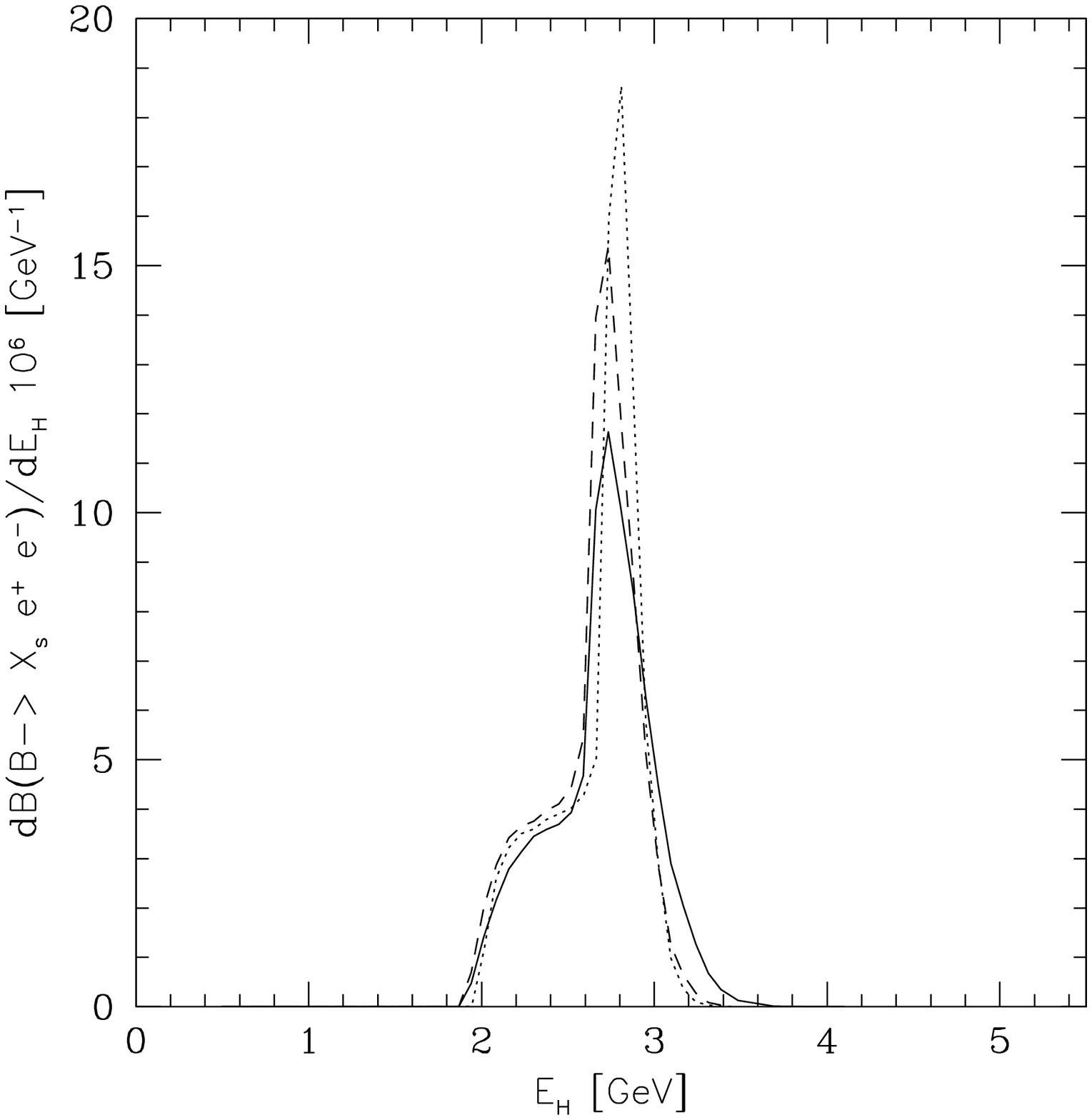,width=7.0cm}
     \end{minipage}
     \hspace{-0.4cm}
     \begin{minipage}[t]{7.0cm}
     \mbox{ }\hfill\hspace{1cm}(d)\hfill\mbox{ }
     \epsfig{file=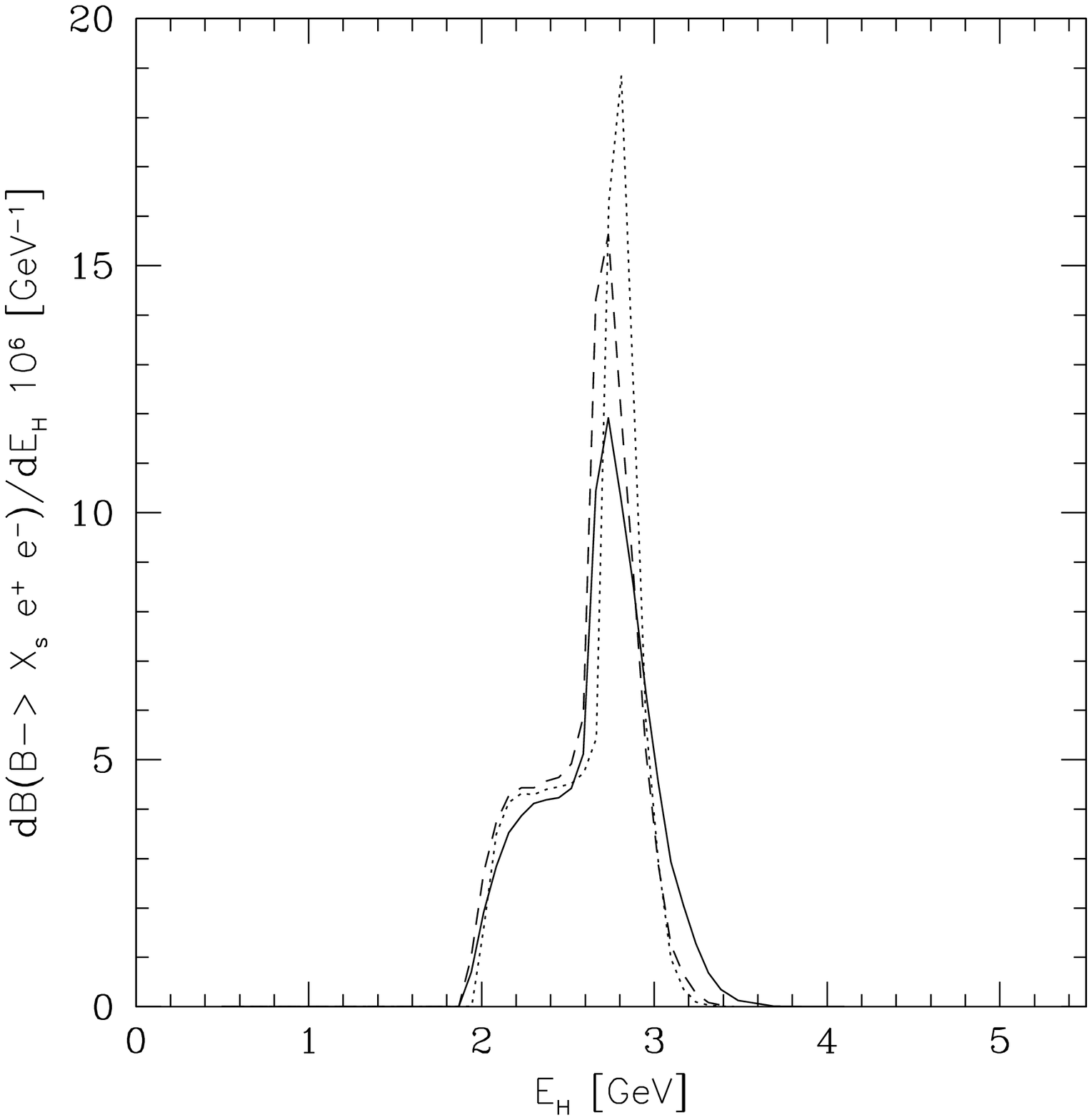,width=7.0cm}
     \end{minipage} \\ 
 \mbox{ }\hspace{-0.7cm}
     \begin{minipage}[t]{7.0cm}
     \mbox{ }\hfill\hspace{1cm}(e)\hfill\mbox{ }
     \epsfig{file=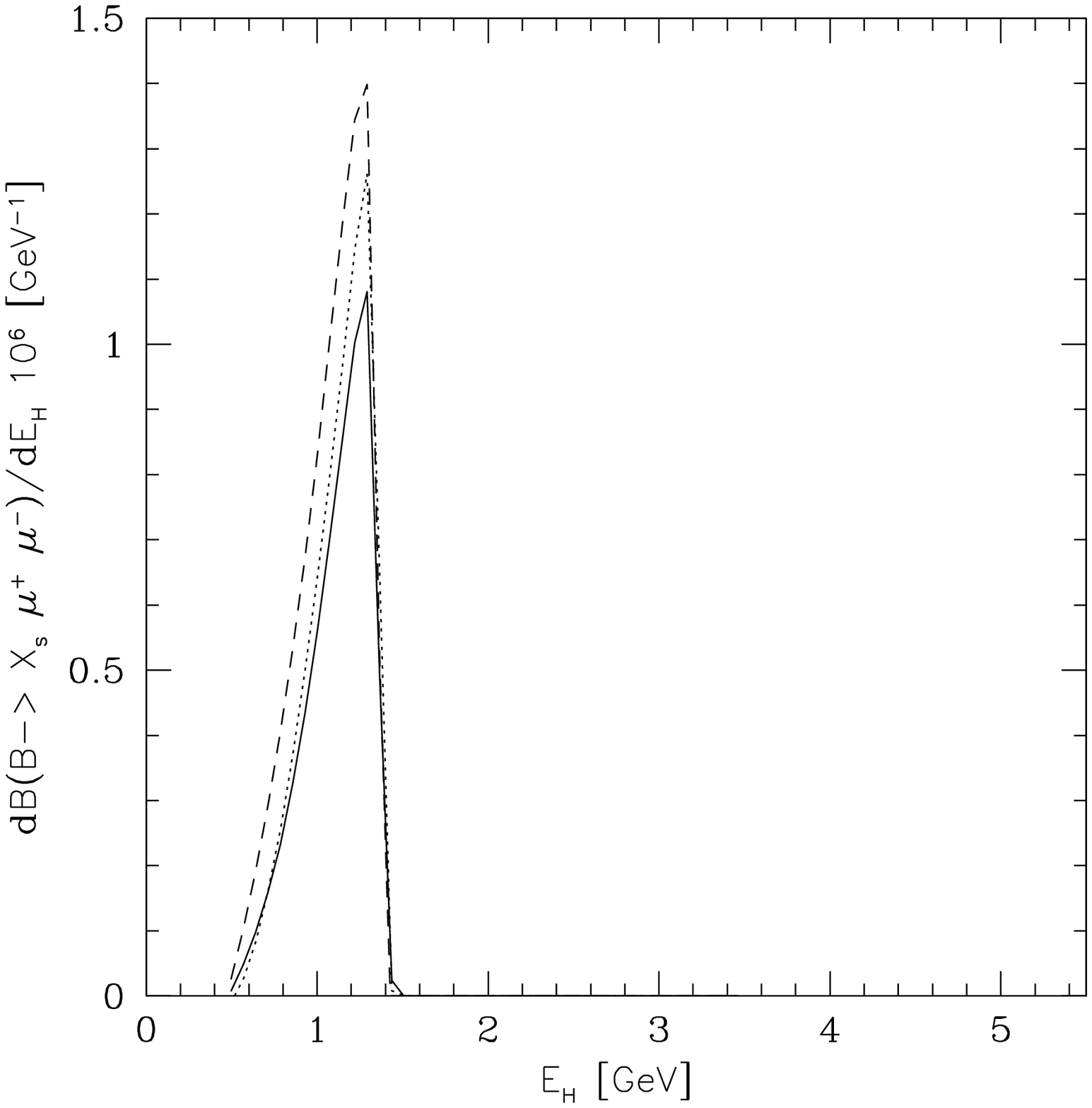,width=7.0cm}
     \end{minipage}
     \hspace{-0.4cm}
     \begin{minipage}[t]{7.0cm}
     \mbox{ }\hfill\hspace{1cm}(f)\hfill\mbox{ }
     \epsfig{file=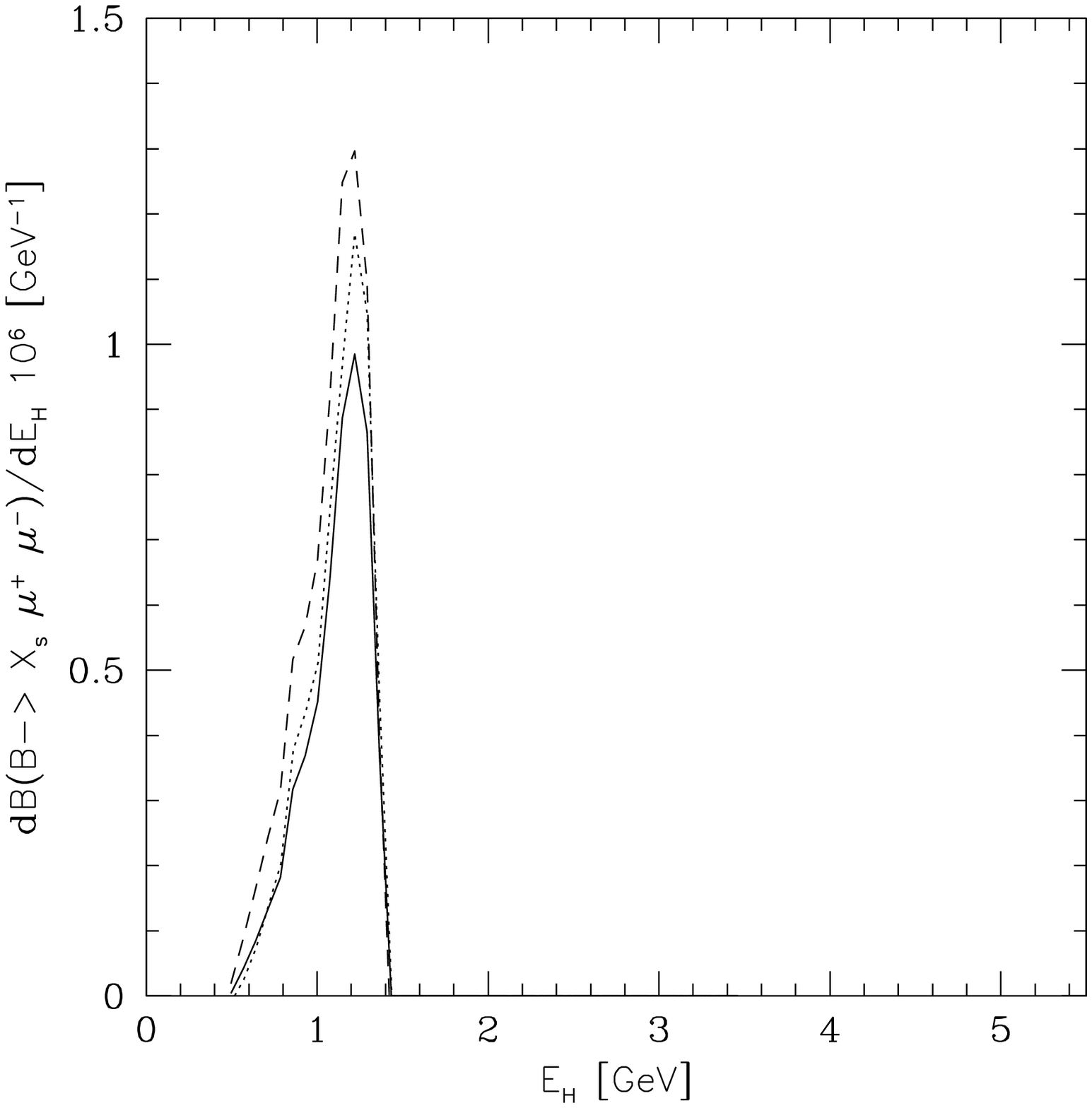,width=7.0cm}
     \end{minipage}
\end{center}  
     \caption{\it 
Hadron energy spectrum in \bxsll in the Fermi motion model with the
cuts on the dilepton mass defined in eq.~(\ref{eq:cuts}); (a),(c),(e) without 
and (b),(d),(f) with the $c\bar{c}$-resonance contribution corresponding to cut
A,B,C, respectively. The  
solid, dotted and dashed curves correspond to the parameters
$ (\lambda_1, \bar{\Lambda})=(-0.3,0.5),(-0.1,0.4),(-0.15,0.35)$ in
(GeV$^2$, GeV), respectively.
}\label{fig:EhLO}
\end{figure}
%
%
%
%
\begin{figure}[t]
\mbox{}\vspace{-2cm}\\
\begin{center}
     \mbox{ }\hspace{-0.7cm}
     \begin{minipage}[t]{7.0cm}
     \mbox{ }\hfill\hspace{1cm}(a)\hfill\mbox{ }
     \epsfig{file=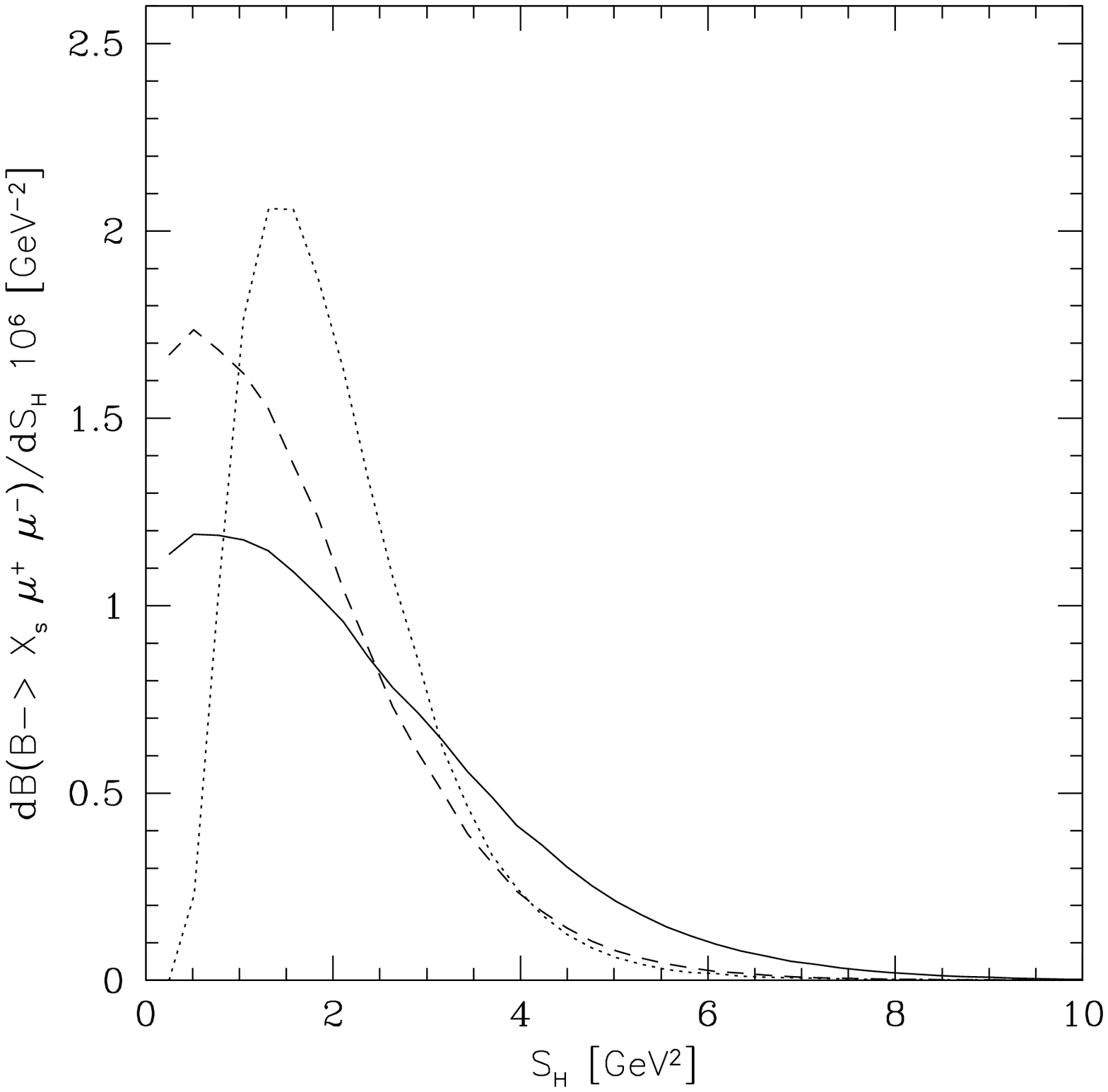,width=7.0cm}
     \end{minipage}
     \hspace{-0.4cm}
     \begin{minipage}[t]{7.0cm}
     \mbox{ }\hfill\hspace{1cm}(b)\hfill\mbox{ }
     \epsfig{file=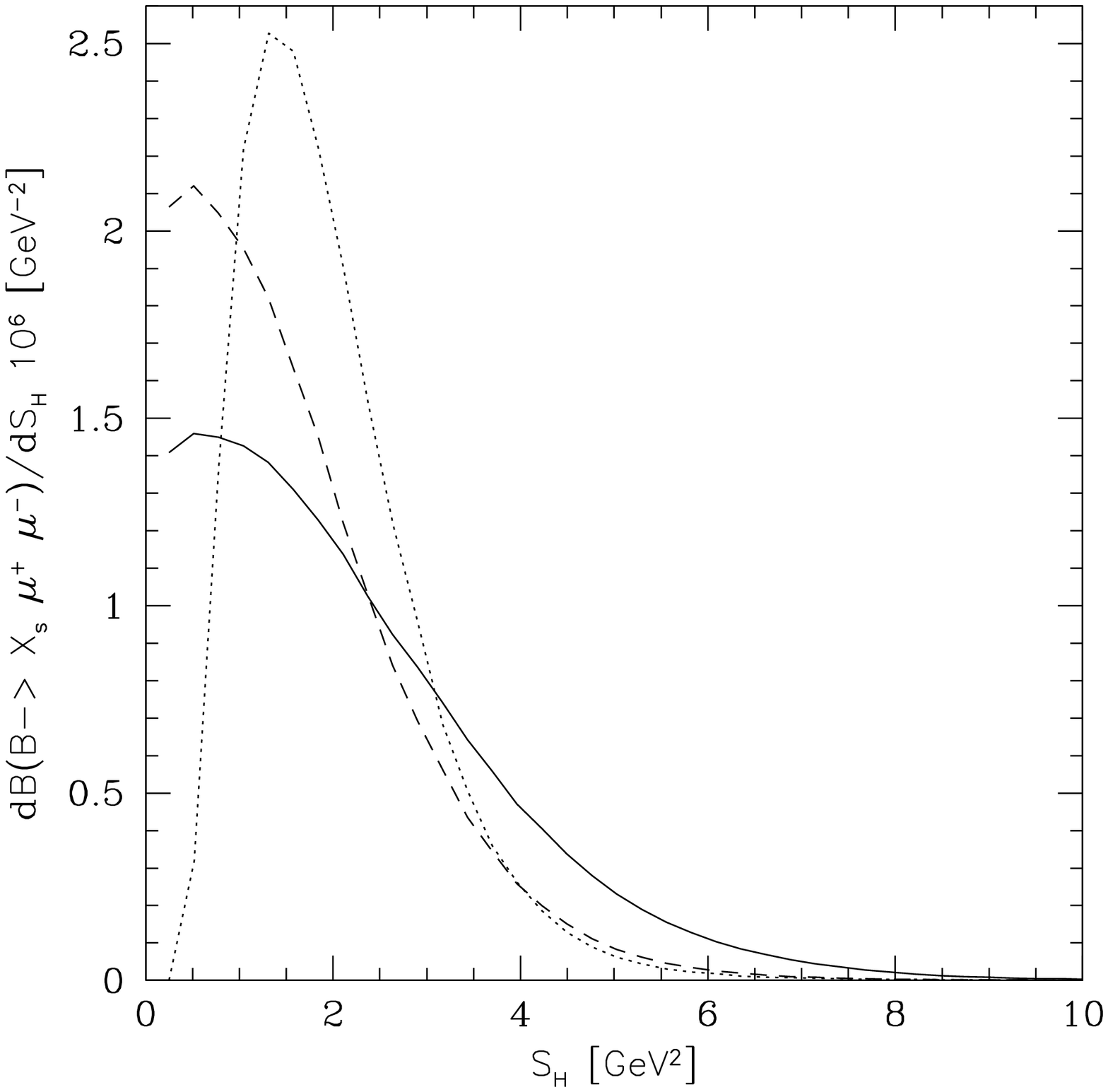,width=7.0cm}
     \end{minipage} \\ 
 \mbox{ }\hspace{-0.7cm}
     \begin{minipage}[t]{7.0cm}
     \mbox{ }\hfill\hspace{1cm}(c)\hfill\mbox{ }
     \epsfig{file=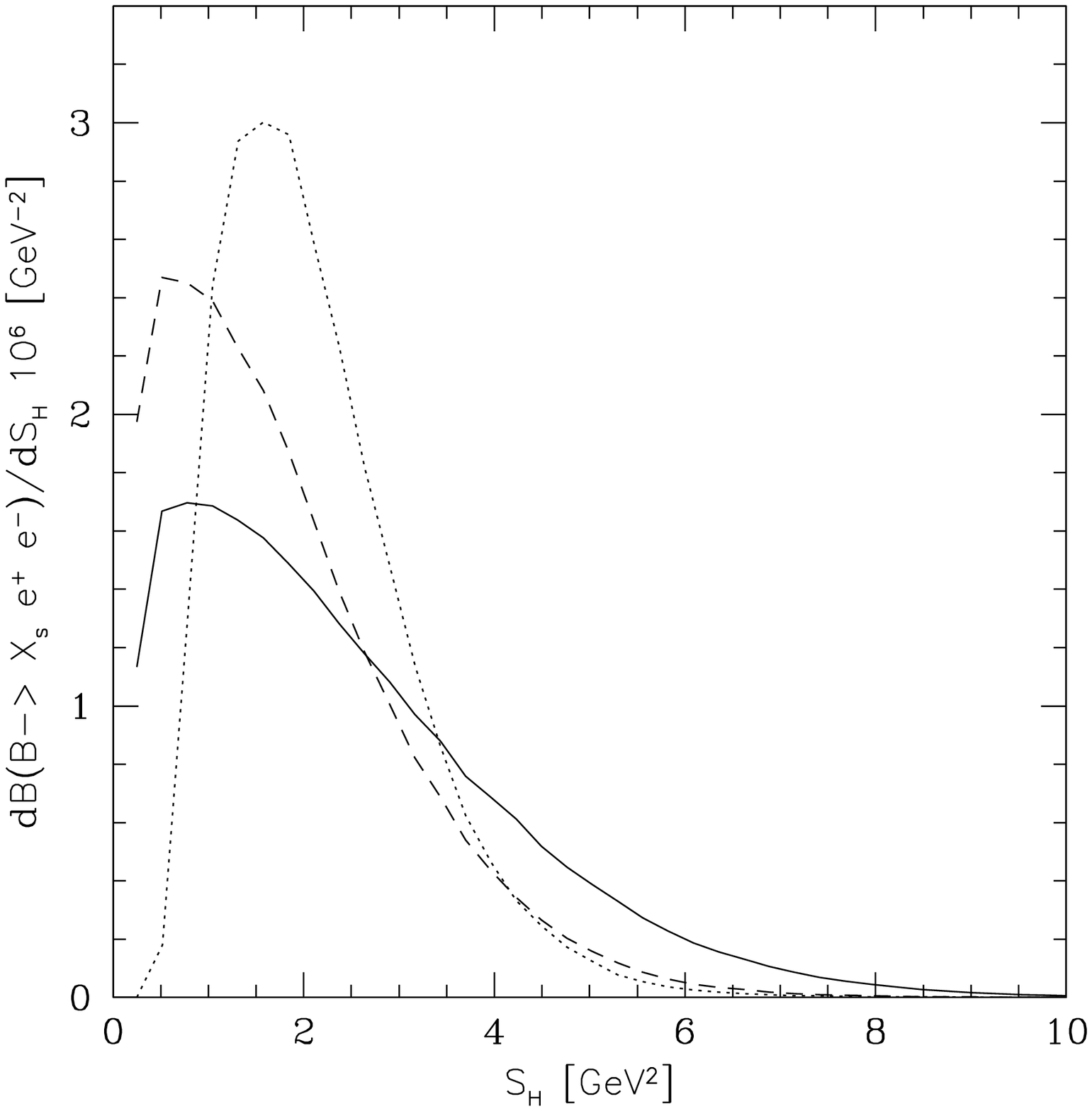,width=7.0cm}
     \end{minipage}
     \hspace{-0.4cm}
     \begin{minipage}[t]{7.0cm}
     \mbox{ }\hfill\hspace{1cm}(d)\hfill\mbox{ }
     \epsfig{file=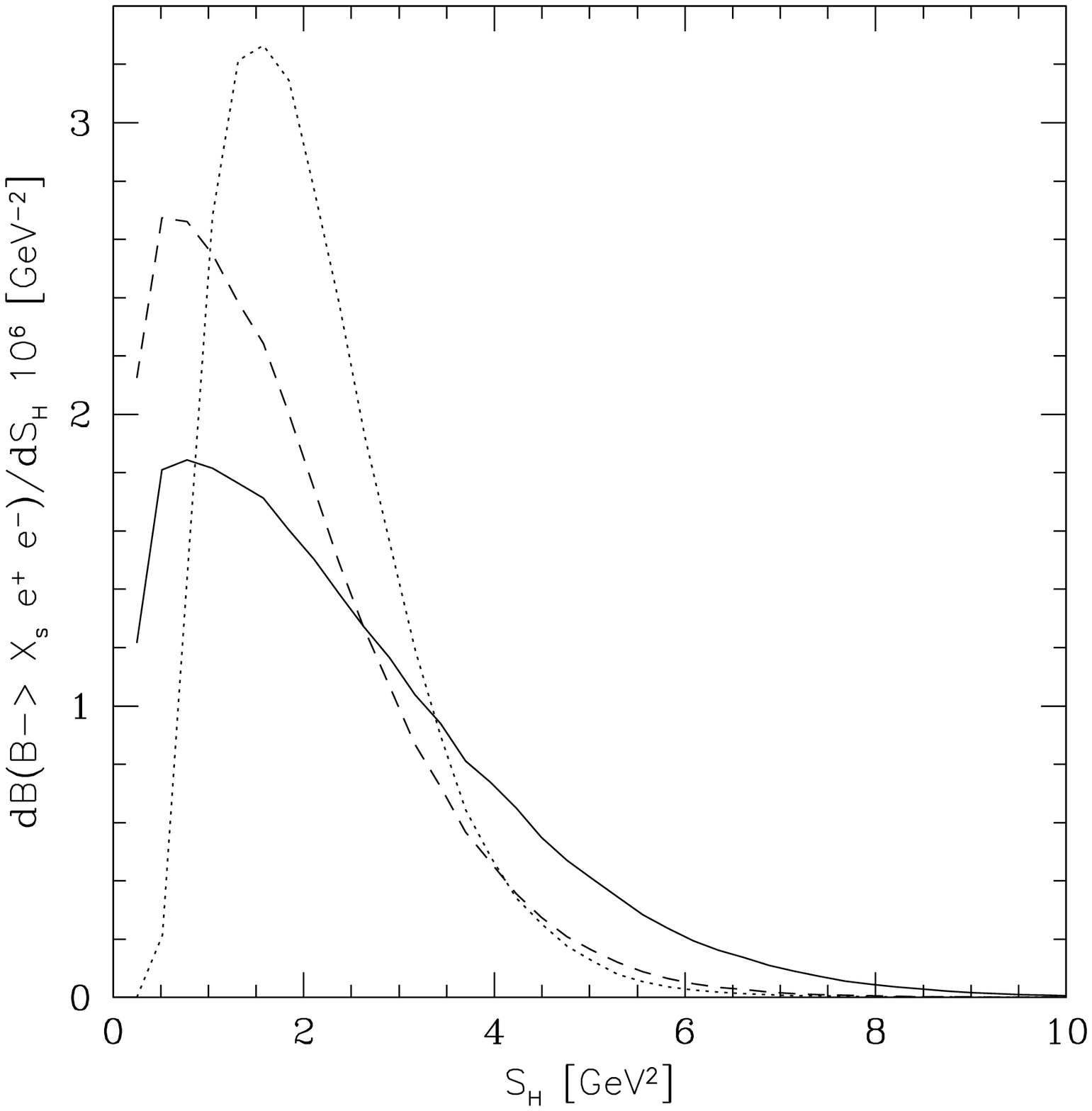,width=7.0cm}
     \end{minipage} \\ 
 \mbox{ }\hspace{-0.7cm}
     \begin{minipage}[t]{7.0cm}
     \mbox{ }\hfill\hspace{1cm}(e)\hfill\mbox{ }
     \epsfig{file=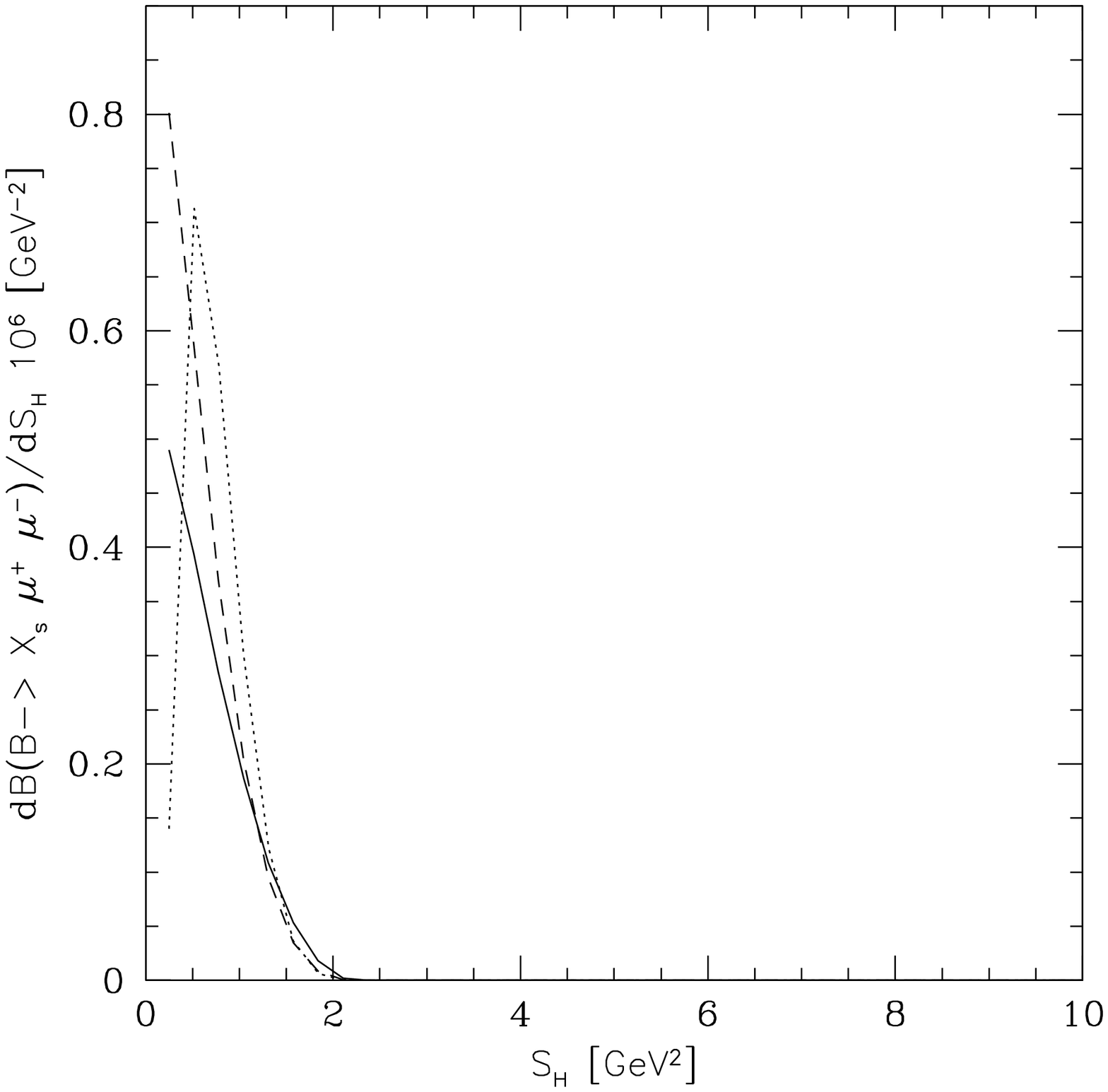,width=7.0cm}
     \end{minipage}
     \hspace{-0.4cm}
     \begin{minipage}[t]{7.0cm}
     \mbox{ }\hfill\hspace{1cm}(f)\hfill\mbox{ }
     \epsfig{file=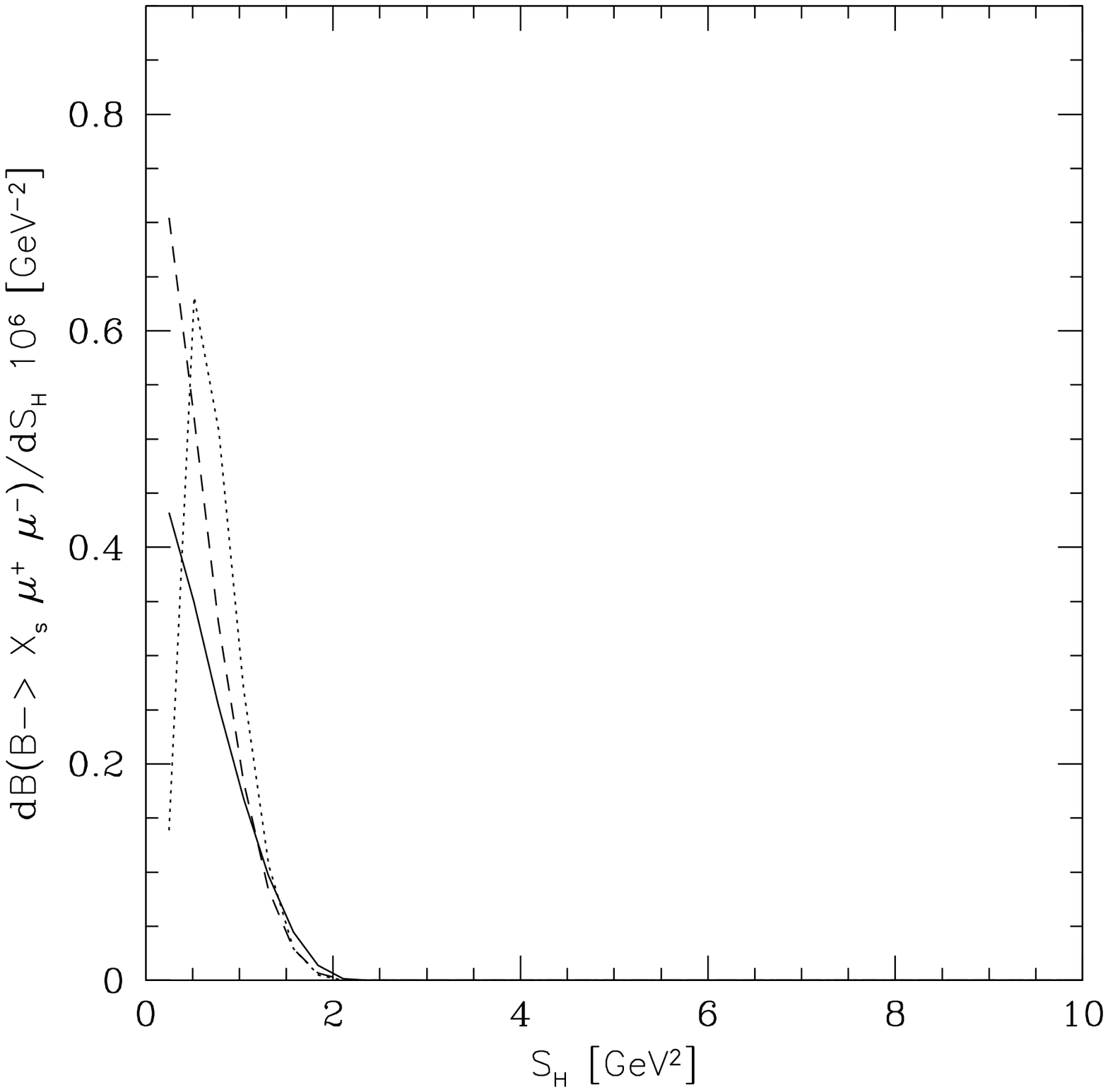,width=7.0cm}
     \end{minipage}
\end{center}  
     \caption{\it 
Hadronic invariant mass spectrum in \bxsll in the Fermi motion model with the
cuts on the dilepton mass defined in eq.~(\ref{eq:cuts}); (a),(c),(e) without 
and (b),(d),(f) with the $c\bar{c}$-resonance contribution corresponding to cut
A,B,C, respectively. The  
solid, dotted and dashed curves correspond to the parameters
$ (\lambda_1, \bar{\Lambda})=(-0.3,0.5),(-0.1,0.4),(-0.15,0.35)$ in
(GeV$^2$, GeV), respectively.
}\label{fig:ShLO}
\end{figure}
\section{Summary and Concluding Remarks}

We summarize our results:
\begin{itemize}
\item We have presented the 
hadron spectra and moments in \bxsll including the resonant-$c\bar{c}$
contribution in the Fermi motion model \cite{aliqcd}.
This complements the description of the final states in \bxsll
presented in \cite{AHHM97}, where the dilepton invariant mass spectrum
and FB asymmetry were worked out in both the HQET and
FM model approaches. We find that the hadron energy spectrum is stable
against variation of the FM model parameters. However, the hadronic
invariant mass is sensitive to the input parameters. This dependence was
already studied for the SD-contribution in
\cite{AH98-1,AH98-2} both in the context of the FM model and HQET.
\item We have quantitatively studied the uncertainties related to the
implementation of the resonant and non-resonant parts in the  
coefficient $C_9^{\mbox{eff}}(\hat{s})$ in \bxsll.
The numerical differences between the approach followed in
\cite{AHHM97} and the alternative ones, discussed in  
\cite{KS96} and \cite{LSW97}, are found to be small in the dilepton 
invariant mass spectrum and negligible in the hadron energy and invariant 
mass spectra and spectral moments. In contrast, theoretical
spectra are found to be more sensitive to the parameters $\lambda_1$ and
$\bar{\Lambda}$.
\item We have studied the hadron spectra by imposing the experimental
cuts designed to suppress the resonant $c\bar{c}$ contributions,
as well as the dominant $B\bar{B}$ background leading to 
the final state $B\bar{B} \to X_s \ell^+ \ell^-$ (+ missing energy).
In particular, the survival probability of the \bxsll signal
resulting from imposing a cut on the hadronic invariant mass 
$S_H < 3.24 \, \mbox{GeV}^2$, as used in the CLEO analysis, 
is estimated and its
parametric dependence studied. We have shown that the cuts such as the ones
used in \cite{cleobsll97} effectively suppress the resonant contribution. 
Thus, the cut spectra essentially test the physics of the short-distance
(and non-resonant $c\bar{c}$) contribution, which can be systematically
studied in perturbation theory and HQET. 
\end{itemize}

 We hope that this work which provides a detailed 
theoretical profile of the hadron spectra in the decay \bxsll will
be helpful in experimental searches of
the rare decay \bxsll. The distributions presented here will allow  
direct comparison of data with SM and will be useful 
in estimating the effects of various experimental cuts on the hadronic
and dilepton invariant masses and hadron energy, which 
will be invoked in experimental analyses. Finally,
this work underscores the importance of 
systematically improving the theoretical precision of the hadron spectra and
spectral moments in the SD-contribution in the decay \bxsll
beyond what has been already done in \cite{AH98-1,AH98-2}, as the 
theoretical uncertainties from the LD-contributions can be brought under 
control by judicious experimental cuts.


\bigskip
\noindent
{\Large \bf Acknowledgements}

We would like to thank Frank Kr\"uger for helpful correspondence
on the implementation of the charmonium contribution given in \cite{KS96}.
We also thank Christoph Greub for useful discussions.
\end{document}